%% file: PrototypeTPC_MuonCalibration.tex
\documentclass[a4paper,11pt]{article}
\pdfoutput=1
\usepackage{jinstpub}              
\usepackage{lineno}

\usepackage{graphicx}
\usepackage{float}
\usepackage{subfig}
\usepackage{url}
\graphicspath{}
\title{Calibration of a Micromegas-based Gaseous Time Projection Chamber Using Cosmic Ray Muons}

\author[a]{Wenming Zhang,}
\author[a]{Yuanchun Liu,}
\author[a]{Haiyan Du,}
\author[a,1]{Ke Han,\note{Corresponding author.}}
\author[a]{Heng Lin,}
\author[e]{Tao Li,}
\author[g]{Lingyin Luo,}
\author[a]{Kaixiang Ni,}
\author[c,d]{Yunzhi Peng,}
\author[a,b,1]{Shaobo Wang,}
\author[c,d]{Sicheng Wen,}
\author[f]{Xiyu Yan,}
\author[c,d]{Zhiyong Zhang,}
\author[a]{Wenchang Zhai}

\affiliation[a]{INPAC and School of Physics and Astronomy, Shanghai Jiao Tong University, MOE Key Lab for Particle Physics, Astrophysics and Cosmology, Shanghai Key Laboratory for Particle Physics and Cosmology, Shanghai 200240, China}
\affiliation[b]{SPEIT (SJTU-Paris Elite Institute of Technology), Shanghai Jiao Tong University, Shanghai 200240, China}
\affiliation[c]{State Key Laboratory of Particle Detection and Electronics, University of Science and Technology of China, Hefei 230026, China}
\affiliation[d]{Department of Modern Physics, University of Science and Technology of China, Hefei 230026, China}
\affiliation[e]{Sino-French Institute of Nuclear Engineering and Technology, Sun Yat-sen University, Zhuhai 519082, China}
\affiliation[f]{School of Physics and Astronomy, Sun Yat-sen University, Zhuhai 519082, China}
\affiliation[g]{School of Physics, Peking University, Beijing 100871, China}

\emailAdd{ke.han@sjtu.edu.cn; shaobo.wang@sjtu.edu.cn}

\abstract{
We report the calibration of a gaseous Time Projection Chamber based on Micromegas charge readout modules with cosmic ray muons, utilizing their penetrating power and relatively uniform energy deposition per unit length.
Muon events were selected through track reconstruction to characterize detector performances, such as the drift velocity, electron lifetime, detector gain, and electric field distortion.
The evolution of detector performances over a 50-day data-taking cycle was measured with the muon calibration method.
For instance, the drift velocity degraded from ${\rm 3.40\pm 0.07 ~ cm/\mu s}$ to ${\rm 3.06\pm 0.06 ~ cm/\mu s}$ without gas purification, and then recovered with gas purification.
A ${\rm ^{137}Cs}$ calibration source was also placed inside the detector as a reference for muon calibrations. 
}


\arxivnumber{}

\begin{document}
\maketitle
\flushbottom

\input{Introduction.tex}
\input{Setup.tex}
\input{Identification.tex}
\input{Results.tex}
\input{Conclusions.tex}

\acknowledgments
This work was supported by grant No.U1965201 and No.11905127 from the National Natural Sciences Foundation of China and grant 2016YFA0400302 from the Ministry of Science and Technology of China. We thank the support from the Key Laboratory for Particle Physics, Astrophysics and Cosmology, Ministry of Education. We thank the support from the Double First Class Plan of Shanghai Jiao Tong University.

\bibliographystyle{unsrt}
\bibliography{bibfile}

\end{document}

%% file: Introduction.tex
\section{Introduction}\label{sec:Introduction}

Time Projection Chamber~(TPC) detectors have been widely used in particle physics since their invention~\cite{Marx:1978zz}. 
Nowadays, TPCs are playing a critical role in rare event search experiments, 
such as Neutrinoless Double Beta Decay~(NLDBD) searches (e.g., NEXT~\cite{NEXT:2015wlq}, EXO-200~\cite{EXO-200:2019rkq}), 
and dark matter direct detections (e.g., XENON1T~\cite{XENON:2018voc}, TREX-DM~\cite{Iguaz:2015myh}, and PandaX-4T~\cite{PandaX-4T:2021bab}). 
A gaseous TPC can record both energy depositions and trajectories of charged particles traveling within, providing information to discriminate the signal from the background~\cite{Gonzalez-Diaz:2017gxo}\cite{Li:2021viv}. 

The PandaX-III~\cite{Chen:2016qcd} is an experiment using a high-pressure xenon gaseous TPC to search for NLDBD of ${\rm ^{136}Xe}$. 
For the prototype TPC of the PandaX-III experiment, we developed a calibration method using cosmic ray muon events to study the key performances, such as the gain, electron lifetime, drift velocity, and electric field distortion. 
Seven Micromegas~(Micro-Mesh Gaseous Structure)~\cite{Giomataris:1995fq} modules are mounted for charge readout in the prototype TPC. 
The Micromegas modules are fabricated with the thermal bonding technology~\cite{Feng:2019prv}, which makes an avalanche structure by pressing the stretched stainless-steel mesh directly onto the readout printed circuit board (PCB) using a hot rolling machine. 
A germanium (Ge) film-based resistive anode is used in Micromegas to suppress the discharge and improve the performance of the detector~\cite{Feng:2022jkd}.
The gas inside the TPC is a conventional gas mixture of argon with 2.5${\rm \%}$ isobutane at 3 bar pressure, which has been proven to perform well in Micromegas testing~\cite{Iguaz:2012ur}.
The muon calibration method can also be applied to study the performance of the detector during the commissioning of 52 Micromegas modules of the PandaX-III experiment, including gain uniformity, electron drift and collection properties, among others.

The TPC records ionization charge signals and measures the relative timing of waveforms to determine the relative position of energy depositions.
However, in a generic gaseous TPC like the PandaX-III, an event's absolute position and time (usually referred to as $t_0$) along the drift direction are unknown without any information from scintillation light signals.
As a result, the drift velocity and electron lifetime cannot be easily measured using conventional calibration methods.
Through-going muon events, which we define as those passing through the readout plane to the cathode plane, have an absolute position since the projected track length along the drift direction is equal to the maximum drift distance of the TPC.
Ideally, cosmic ray muons have long-straight tracks and a near-constant ionization energy loss per unit length dE/dx~\cite{Groom:2001kq}. 
Using these characteristics, many cosmic ray tests have been performed in the gas and liquid TPCs~\cite{Kobayashi:2010hx}\cite{WA105:2021zin}.
These properties of muons allow us to calibrate the key performances of the TPC conveniently and efficiently. 

%% file: Setup.tex
\section{Experimental setup and data taking}\label{sec:Setup}

\begin{figure}[tbp]
    \centering
    \includegraphics[width=.75\textwidth]{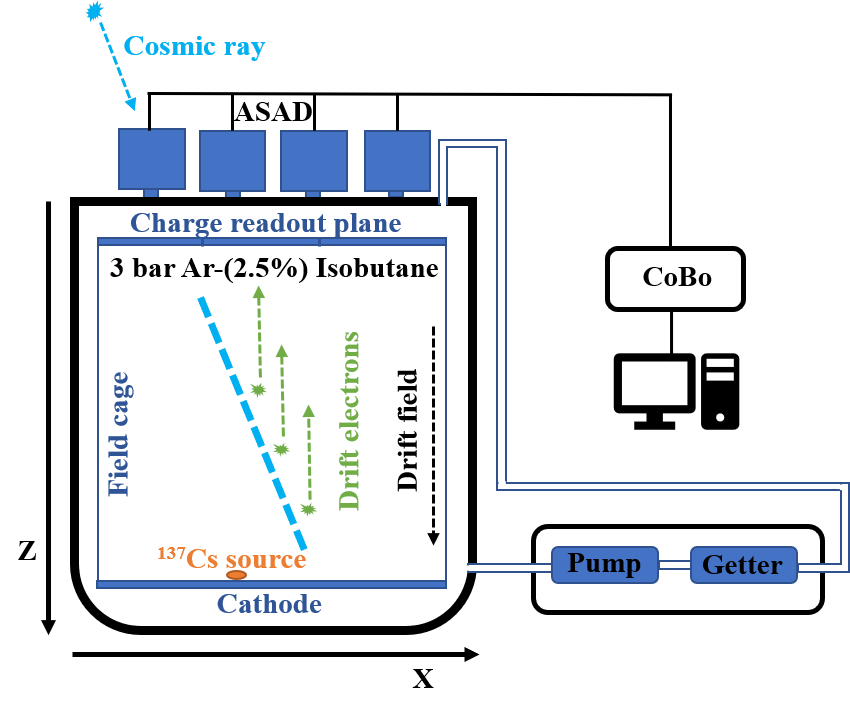}
    \caption{\label{fig:prototype_sketch}  Structure of the TPC with the gas system and electronics. The charge readout plane is defined as the horizontal XY plane, with the drift field pointing towards the vertical Z direction. The maximum drift distance of the detector is 78.0 cm. }
\end{figure}

\begin{figure}[tbp]
    \centering
    \includegraphics[width=.60\textwidth]{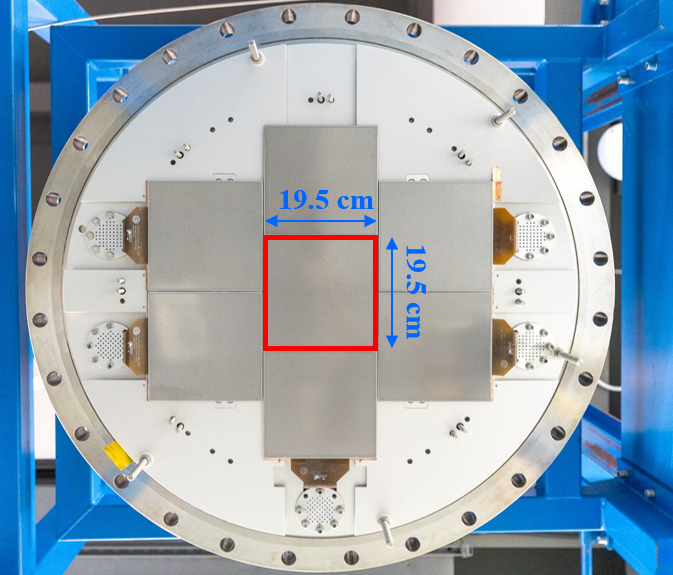}
    \includegraphics[width=.343\textwidth]{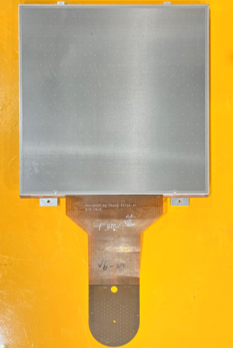}
    \caption{\label{fig:readout_plane} (Left) Picture of the charge readout plane during assembly. The location of the Micromegas for data taking is in the red box. (Right) Picture of one 19.5${\rm \times}$19.5 ${\rm cm^2}$ thermal bonding Micromegas module.}
\end{figure}

The core component of the experimental setup is a gaseous TPC, which consists of a charge readout plane on the top, a cathode at the bottom, and an electric field shaping cage in the middle. 
Fig.~\ref{fig:prototype_sketch} demonstrates the schematic of the TPC and sub-systems, including a gas system and readout electronics. 
The TPC vessel has a total inner volume of about 600 L.
The active volume inside the TPC is a cylinder with a diameter of 66.0 cm and a height of 78.0 cm (about 270 L), surrounded by the electric field shaping cage made of an acrylic barrel and flexible PCBs~\cite{Wang:2020owr}. 
More description of the design and construction of the TPC can be found in~\cite{Lin:2018mpd}. 
As shown in Fig.~\ref{fig:readout_plane}, the charge readout plane consists of a tessellation of seven 19.5${\rm \times}$19.5 ${\rm cm^2}$ thermal bonding Micromegas modules.
Each module has 64 readout strips of 3~mm pitch in each direction (X or Y).
All 128 strips are set in the inner layer of the readout PCB, with the upper layer oriented in the X direction and the lower layer in the Y direction, and then connected to the electronics via flexible signal lines.
The spacers made of the thermal bonding film are used to support the stainless-steel mesh onto the readout anode, creating an amplification gap of 100 ${\rm \mu m}$.
In order to study the detector calibration method, only the center Micromegas module was activated for data taking in this work.
The relevant active volume for single-module measurements is about 30 L.
A ${\rm ^{137}Cs}$ calibration source, wrapped in Kapton to shield the beta particles from escape, was placed on the cathode plane under the active Micromegas. 

When an incident particle travels in the active volume, it deposits energy by ionizing the gas atoms along its trajectory. 
With a negative high voltage applied to the cathode, the primary ionization electrons drift along the electric field and get collected by the Micromegas strips in X and Y directions after avalanche amplification. 
The strip signals are digitized by a commercial ASAD (ASIC Support and Analog-Digital conversion) and CoBo (Concentration Board) electronics system~\cite{Giovinazzo:2016ikh} with a record length of 512 sampling points. 
A CoBo card reads up to 4 ASAD boards, and an ASAD board hosts 4 AGET (ASIC for Generic Electronic system for TPCs) chips~\cite{Anvar2011AGETTG}, each of which can process 64 channels from the detector input.
The chip has a sampling frequency up to 100 MHz and a dynamic range from 120 fC to 10 pC. 
To account for the maximum drift time of through-going muon events, a sampling frequency of 10 MHz~(a time window of 51.2 ${\rm \mu s}$) is used for data taking.
The strip signals provide the detected charges and the horizontal positions in the XY plane. 
The timing information of signals provides a relative position measurement in the Z direction. 
Therefore, a complete three-dimensional track and the energy deposition of an event are recorded. 

A gas system was constructed for gas circulation and purification, as shown in Fig.~\ref{fig:prototype_sketch}.
Gas impurities, especially electronegative ones, absorb the drift electrons, resulting in a so-called attachment effect that degrades the detector performance.
As our detector is hermetic, the main contribution of gas impurities is from the outgassing inside the detector~\cite{NEXT:2018wtg}\cite{Lin:2022kua}. 
Continuous purification with a circulation loop is necessary to reduce contamination levels. 
A room-temperature getter, model HP190-702FV by SAES~\cite{SAES}, is used to remove gas impurities such as ${\rm H_{2}O}$, ${\rm O_{2}}$, CO, and ${\rm CO_{2}}$ effectively. 
A magnetically driven circulation pump drives the gas mixture through the getter at a speed of 3 standard liters per minute. 

\begin{figure}[tbp]
    \centering
    \includegraphics[width=.47\textwidth]{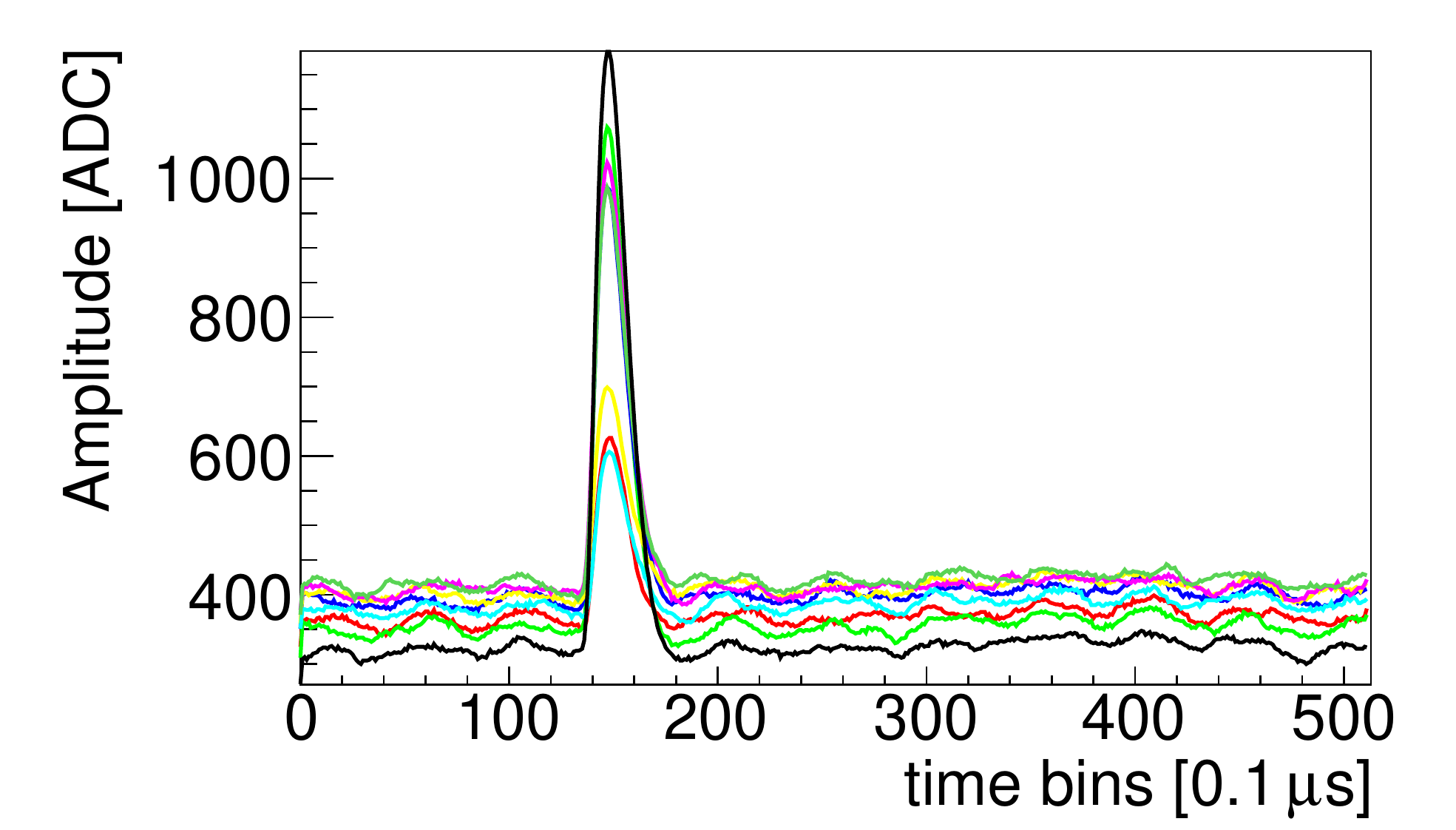}
    \qquad
    \includegraphics[width=.47\textwidth]{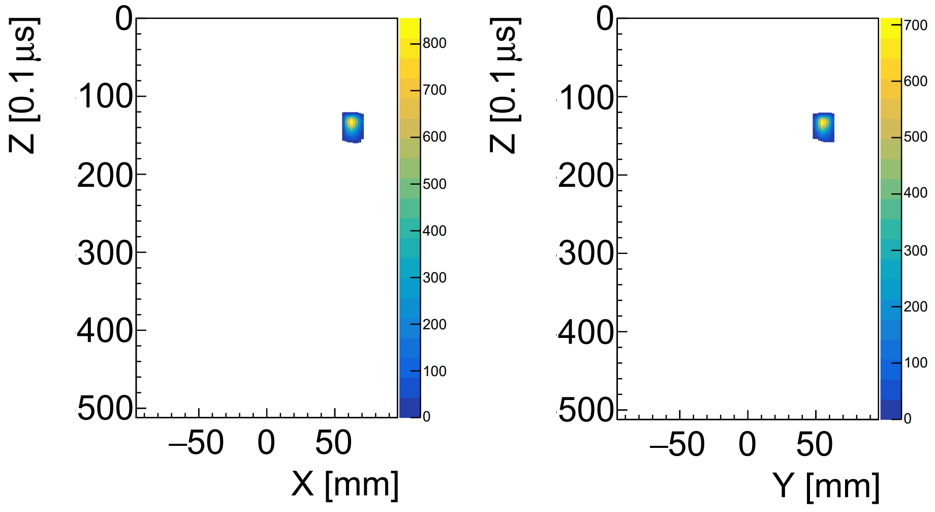}
    \qquad
    \includegraphics[width=.47\textwidth]{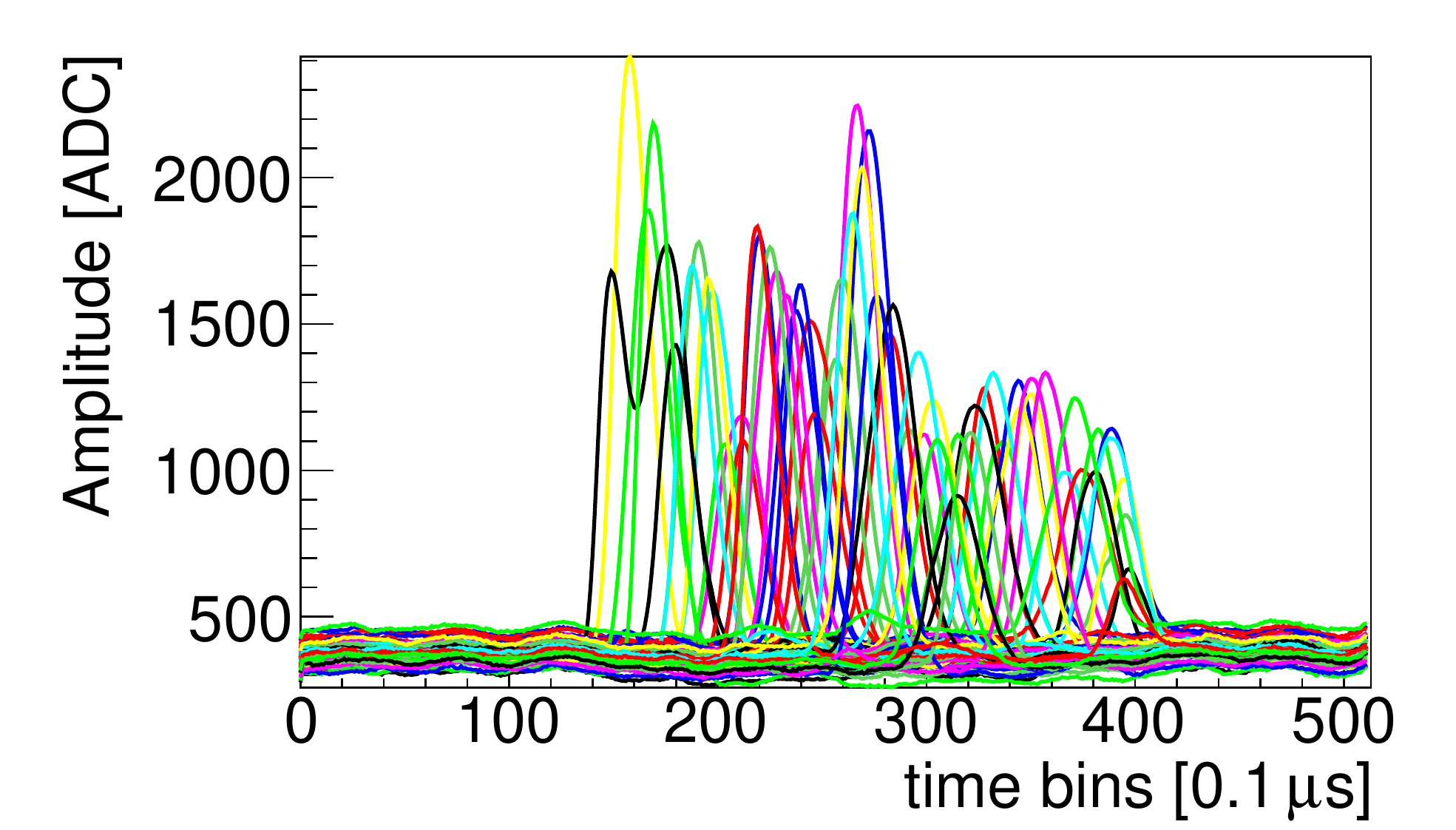}
    \qquad
    \includegraphics[width=.47\textwidth]{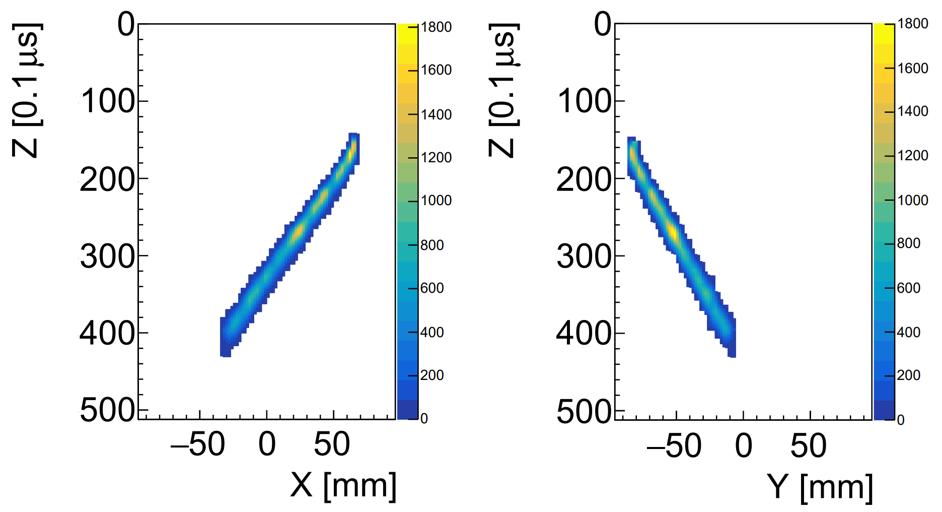}
    \caption{\label{fig:pulse_and_track}  (Top) An example of a characteristic X-ray event from the ${\rm ^{137}Cs}$ source. (Bottom) An example of a through-going muon event from cosmic rays. Signal pulses collected on each triggered strip are shown in the left panel, where one color indicates one channel. The projected tracks on the XZ and YZ planes are shown in the right panel.}
\end{figure}

We took data for 50 days with the same detector configuration, including 48 days without gas purification and two days with gas purification, to study the TPC performance evolution.
The detector was filled with 3 bar Ar-(2.5${\rm \%}$)Isobutane on Oct. 25, 2022. 
We applied 520 V for the Micromegas mesh and 24 kV for the cathode, corresponding to an amplification field of 52 kV/cm and a drift field of 301 V/cm.  
Additionally, we studied the electron drifting process and electric field distortion with different electric fields in the same gas condition.
A ${\rm ^{137}Cs}$ source placed on the cathode was used for reference.

Events from both cosmic ray muons and the ${\rm ^{137}Cs}$ source were selected for detector calibration. 
Signals collected in X and Y strips of Micromegas allow us to obtain projected tracks on the XZ and YZ planes. 
The total energy of an event is determined by summing the collected charges on strips. 
Two example events, including their tracks and waveforms, are shown in Fig.~\ref{fig:pulse_and_track}. 
Events from the ${\rm ^{137}Cs}$ source only triggered an average of eight strips, corresponding to its characteristic X-ray of 32.2 keV. 
However, muon events exhibit distinctive long-straight tracking features. 
We describe the muon event identification through track reconstruction in the next section. 

%% file: Identification.tex
\section{Muon event identification}\label{sec:Identification}

\begin{figure}[tbp]
    \centering
    \includegraphics[width=.23\textwidth]{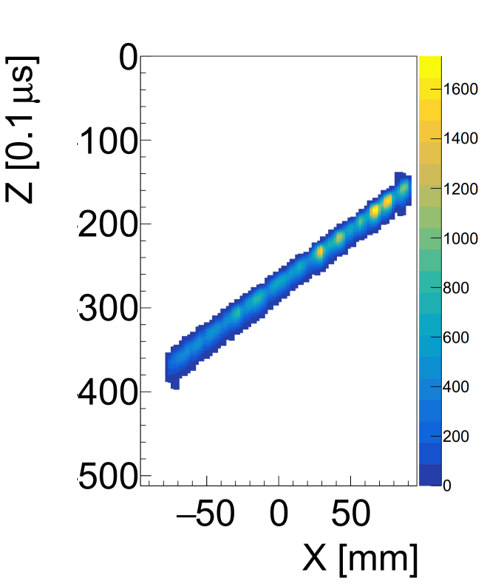}
    \includegraphics[width=.23\textwidth]{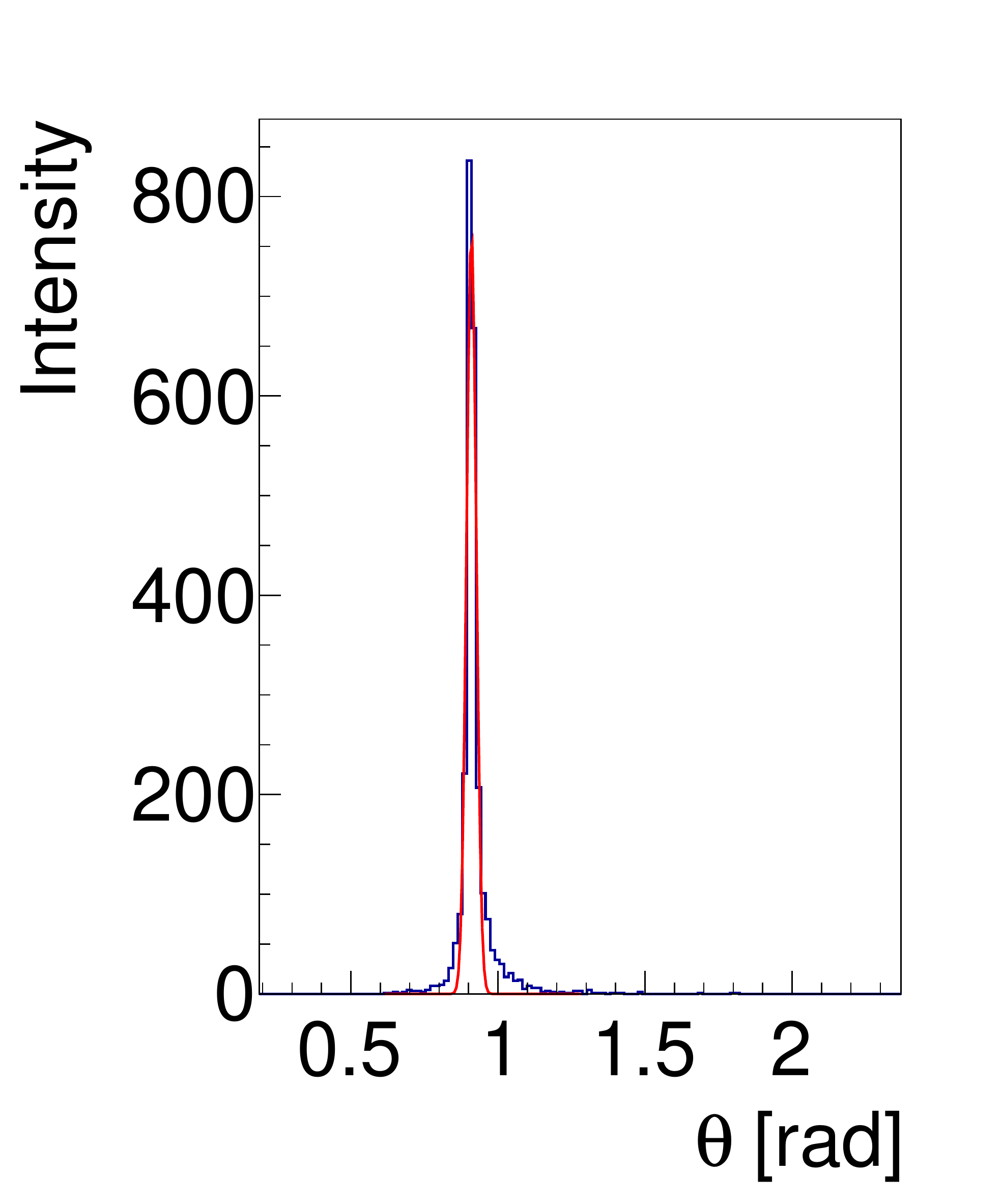}
    \qquad
    \includegraphics[width=.23\textwidth]{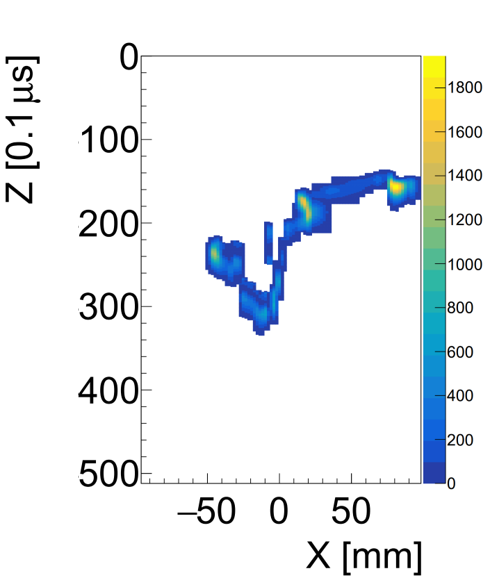}
    \includegraphics[width=.23\textwidth]{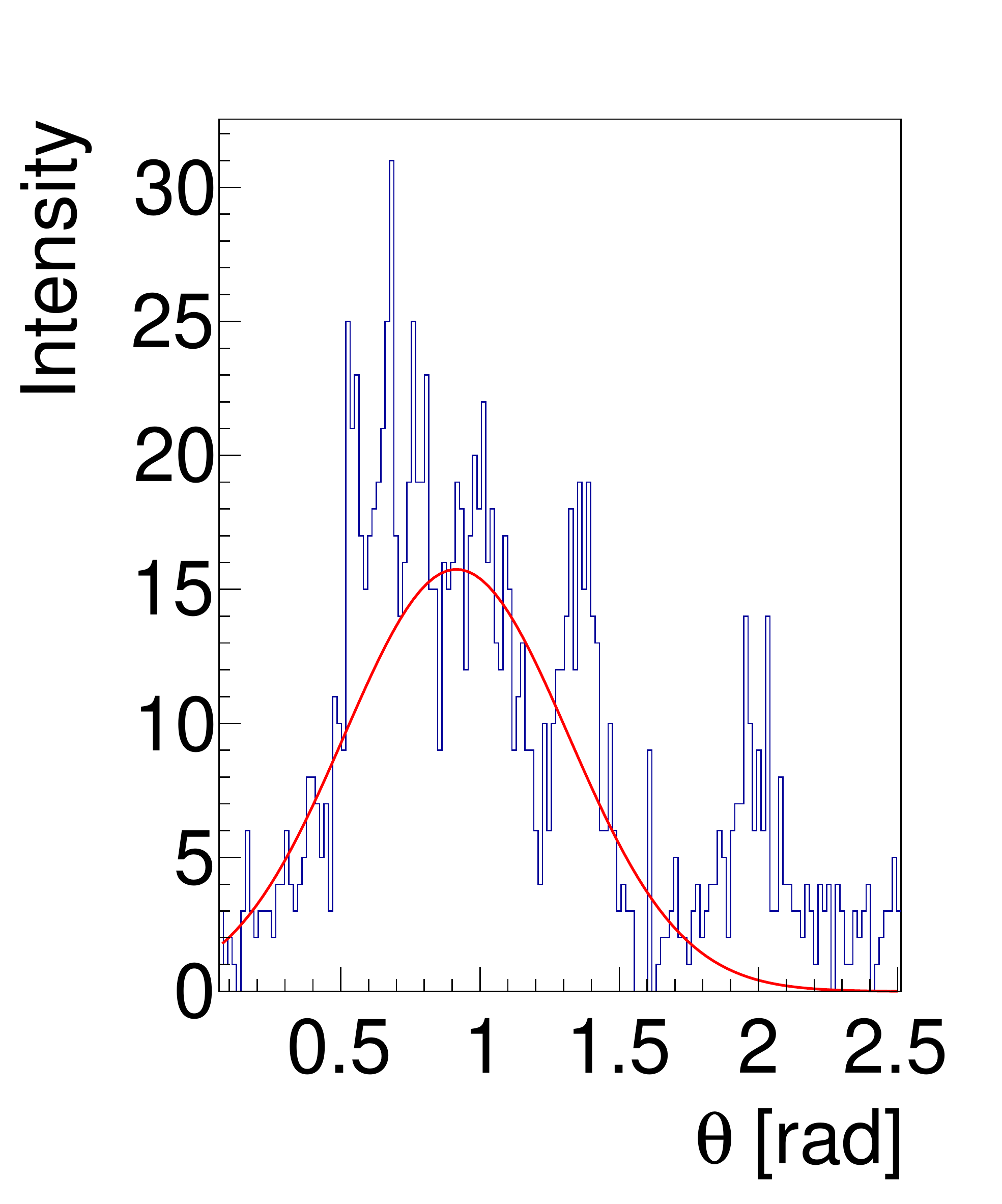}
    \caption{\label{fig:hough_theta_and_tracks}  (Left) An example of a muon track and its ${\theta}$ distribution from the Hough transform. (Right) An example of a typical high-energy electron track and its ${\theta}$ distribution from the Hough transform.}
\end{figure}

Muon and ${\rm ^{137}Cs}$ X-ray events can be identified based on their track lengths and distributions in the detector. 
However, both muon and high-energy electron events have long tracks in the TPC with dozens of strips triggered.
We used the Hough transform algorithm~\cite{Duda:1972ymn} to reconstruct the tracks and calculate the degree of the track's zigzag to discriminate muons from electrons.
The projected tracks were discretized into many track points in the XZ (or YZ) plane. 
The line between any two track points can be described by an angle ${\theta}$ and a radius $\rho$ under the polar coordinate system. 
As shown in Fig.~\ref{fig:hough_theta_and_tracks}, the fitted Gaussian sigma of the ${\theta}$ distribution distinguishes muons from high-energy electrons straightforwardly. 
The ${\theta}$ distribution of a straight muon track is narrow, while a zigzagging high-energy electron track gives a broad ${\theta}$ distribution due to multiple scattering. 
The mode of the ${\theta}$ distribution was used to represent the expected direction of a straight muon track.
Track reconstruction and identification were performed in the REST framework~\cite{Altenmuller:2021slh}.

\begin{figure}[tbp]
    \centering
    \subfloat[]{
        \label{subfig:muon_hitmap}
        \includegraphics[width=.45\textwidth]{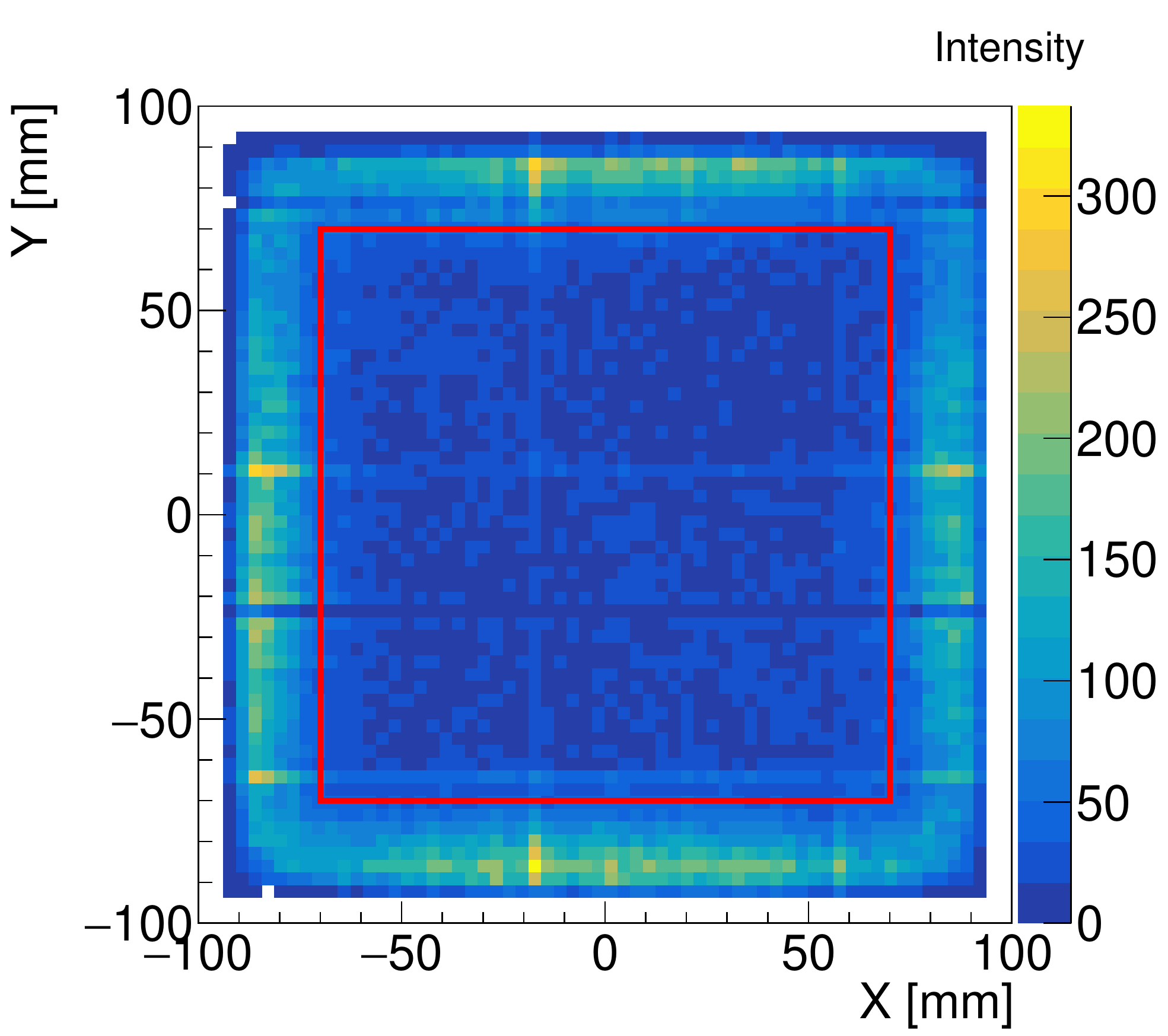}
    }
    \subfloat[]{
        \label{subfig:source_hitmap}
        \includegraphics[width=.45\textwidth]{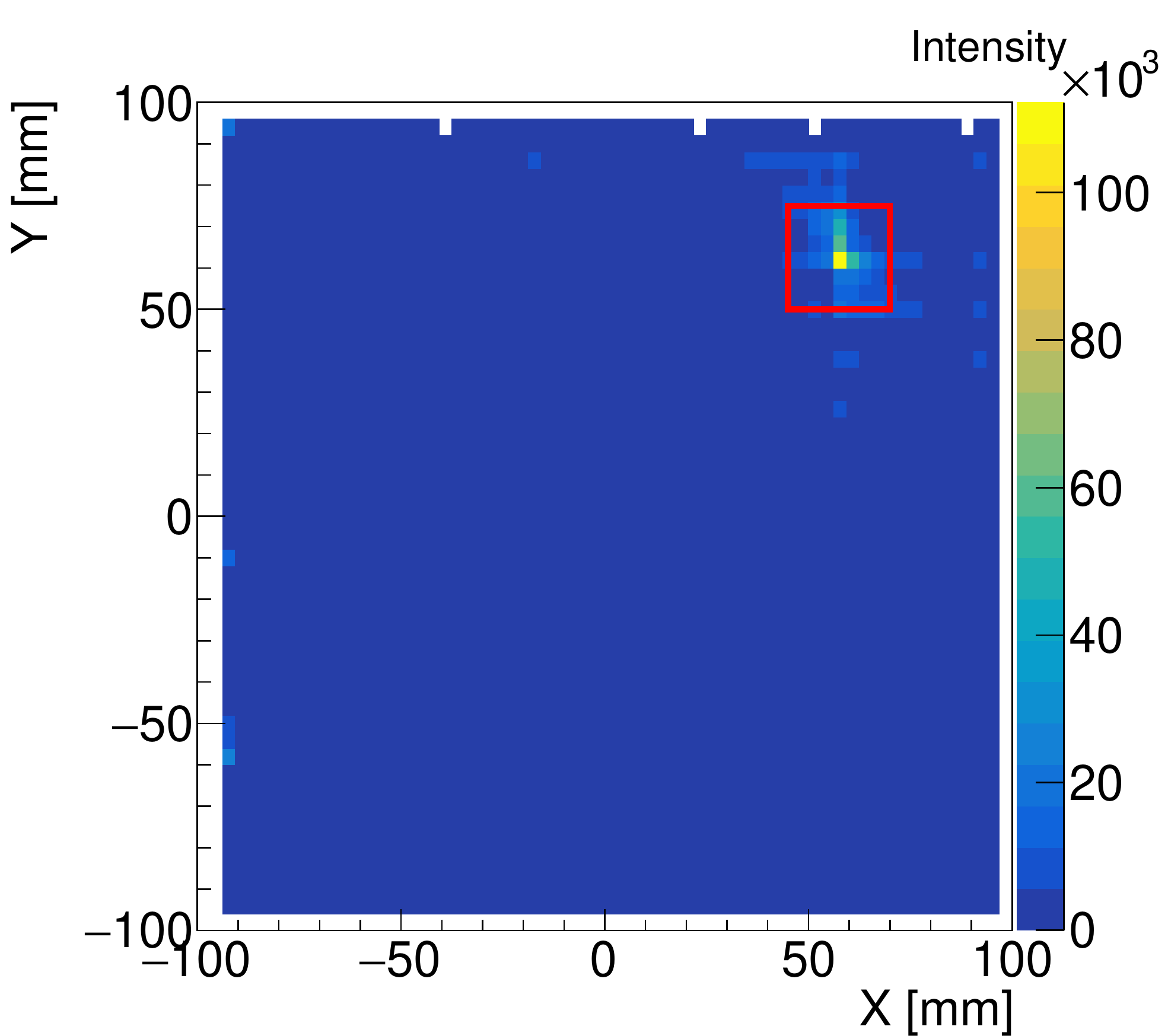}
    }
    \caption{\label{fig:hitmap} (a) A hit map of the first triggered position of cosmic-ray muons. Signals in the red square were selected by the positioning cut for detector calibrations. (b) A hit map of the last triggered position of characteristic X-ray events from the ${\rm ^{137}Cs}$ source. The hot spot in the red box is the image of the ${\rm ^{137}Cs}$ source.}
\end{figure}

After track identification, we selected through-going muons by the positioning cut and drift time distribution cut for detector calibrations. 
The through-going muon goes through the top and bottom of the TPC active volume, and the projected track falls within the Micromegas module.  
Its track starts from the readout plane, and the first triggered Z coordinate equals zero. 
For single-module measurements, a rectangular active volume is shaped with a Micromegas module of 19.5${\rm \times}$19.5 ${\rm cm^2}$ and a drift distance of 78.0 cm. 
As a result, many muons enter from the side of the active volume, first triggering the boundary strips and forming the highest intensity near the boundary, as shown in Fig.~\ref{subfig:muon_hitmap}.
A positioning cut within the red square of ${\rm 140 \times 140~mm^2}$ on the hit map, was used to exclude side-entering muons. 
With the absolute Z position and detected charges of each track point, through-going muon events can be used to characterize the electron lifetime and detector gain. 
Furthermore, the electric field distortion can be measured through the track deviation of muon events.

Events from the ${\rm ^{137}Cs}$ source are mainly distributed near the source position and leave short tracks in the TPC volume. 
The hit map presented in Fig.~\ref{subfig:source_hitmap} is generated from the last triggered-strip position of events, and the highlighted cluster reveals the expected position of the ${\rm ^{137}Cs}$ source. 
We selected source signals in the red square of ${\rm 25 \times 25~mm^2}$ by the positioning cut and used them as a reference to calibrate the detector gain and energy resolution.

%% file: Results.tex
\section{Detector performance results}\label{sec:Results}
\subsection{Drift velocity and electron lifetime}

In a TPC, ionization electrons drift to the readout plane under the influence of the drift field.
The drift velocity ${v_{drift}}$ depends on the gas mixture and electric drift field.
A portion of ionization electrons is absorbed by electronegative impurities during drifting, resulting in the attachment effect characterized by the electron lifetime ${\tau_e}$ and electron absorption distance ${z_e}$~\cite{Huxley:1974}. 
The effect can be expressed as an exponential function:
\begin{equation}
    \label{eq:electron_lifetime_calculation}
    E = E_0 e^{-z/z_e} = E_0 e^{-t/{\tau}_e},
\end{equation}
where ${E}$ is the energy detected by the readout plane, ${E_0}$ is the energy deposited in the TPC, and ${z}$ (${t}$) is the drift distance (time) of ionization electrons. 
By combining the calculated drift velocity, we can estimate the electron lifetime (${\tau_e = z_e / v_{drift}}$). 
The drift velocity and electron lifetime measurement are crucial to understanding the drift process in the TPC. 

\begin{figure}[tbp]
    \centering
    \includegraphics[width=0.6\textwidth]{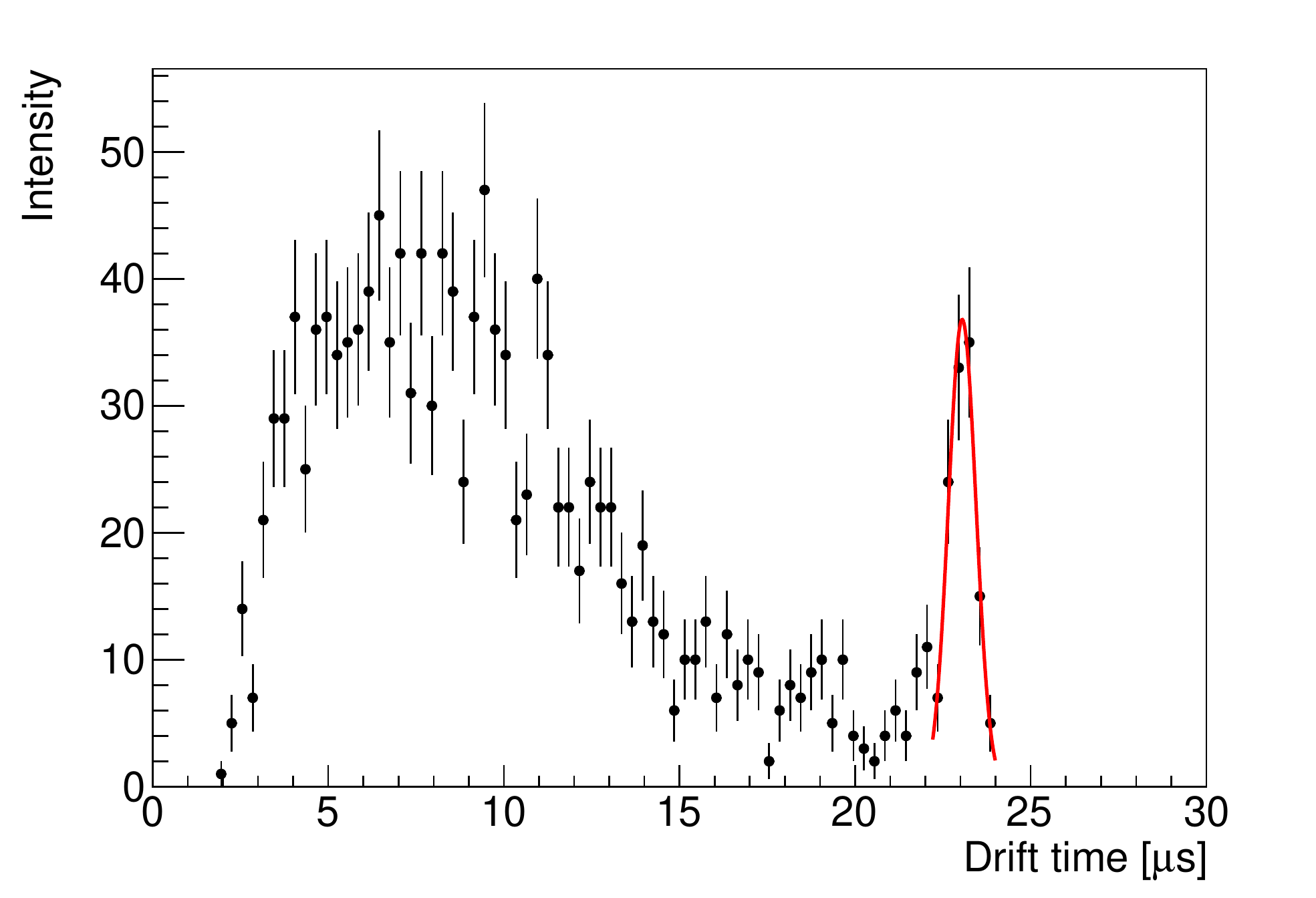}
    \caption{\label{fig:drift_time} The drift time distribution of selected muon events. The peak at the tail of the distribution fitted with a Gaussian function represents through-going muons.}
\end{figure}

\begin{figure}[tbp]
    \centering
    \includegraphics[width=0.9\textwidth]{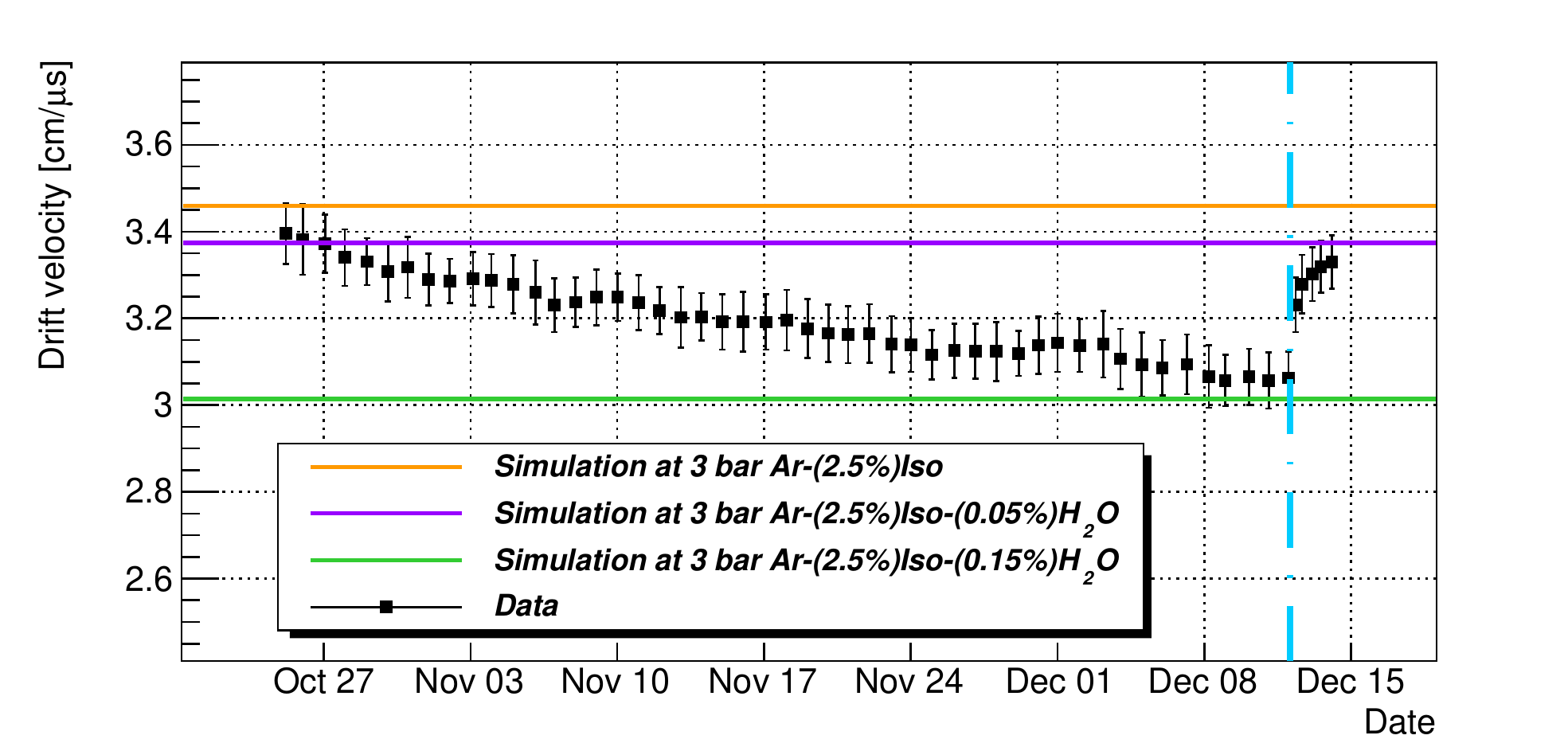}
    \caption{\label{fig:driftVelocity_vs_time} The drift velocity of ionization electrons evolves with time. Three colored lines with different gas mixtures from Garfield++ are also shown in the figure. The blue dashed line denotes the starting time of the circulation and purification of the gas system. Data were taken at 3 bar Ar-(2.5\%)isobutane gas mixture and a drift field of 301 V/cm.}
\end{figure}

Through-going muons can be used to calculate the drift velocity and quantify the electron absorption during drifting. 
Fig.~\ref{fig:drift_time} shows the drift time distribution of muons on the first day of data taking.
The peak was fitted with a Gaussian function, whose fitted mean was used to calculate the maximum drift time of through-going muons, and sigma was used to calculate the uncertainties.
A maximum drift time of ${\rm 22.9\pm 0.5 ~ \mu s}$ is obtained with a maximum drift distance of 78.0~cm.
The corresponding drift velocity is ${\rm 3.40\pm 0.07 ~ cm/\mu s}$ at 3 bar Ar-(2.5\%)isobutane gas mixture and a drift field of 301 V/cm. 
Our initial drift velocity is in fair agreement with the Garfield++ calculated value of ${\rm 3.46 ~ cm/\mu s}$~\cite{Garfield++}.
Fig.~\ref{fig:driftVelocity_vs_time} shows the evolution of drift velocity in the data-taking process. 
The drift velocity decreased to ${\rm 3.06\pm 0.06 ~ cm/\mu s}$ after one and a half months, due to the deterioration of gas quality caused by outgassing in the detector.
When the gas circulation and purification were turned on, the drift velocity quickly recovered in two days. 
Outgassing from detector components, such as oxygen and water, can significantly impact performance.
For example, acrylic that makes up the field cage tends to trap water easily, with a water absorption ratio up to ${\rm 0.4\%}$ by weight~\cite{matweb}.
Profuse outgassing from acrylic may introduce significant water contaminants to the detector~\cite{NEXT:2018wtg}.
As the contamination level inside the TPC is difficult to measure precisely, the figure shows three simulation results with different gas mixtures from Garfield++. 
If 0.05\% (0.15\%) water contaminants are added to the gas mixture, the simulated drift velocity decreases due to the increased electron scattering cross-section. 

\begin{figure}[tbp]
    \centering
    \includegraphics[width=0.9\textwidth]{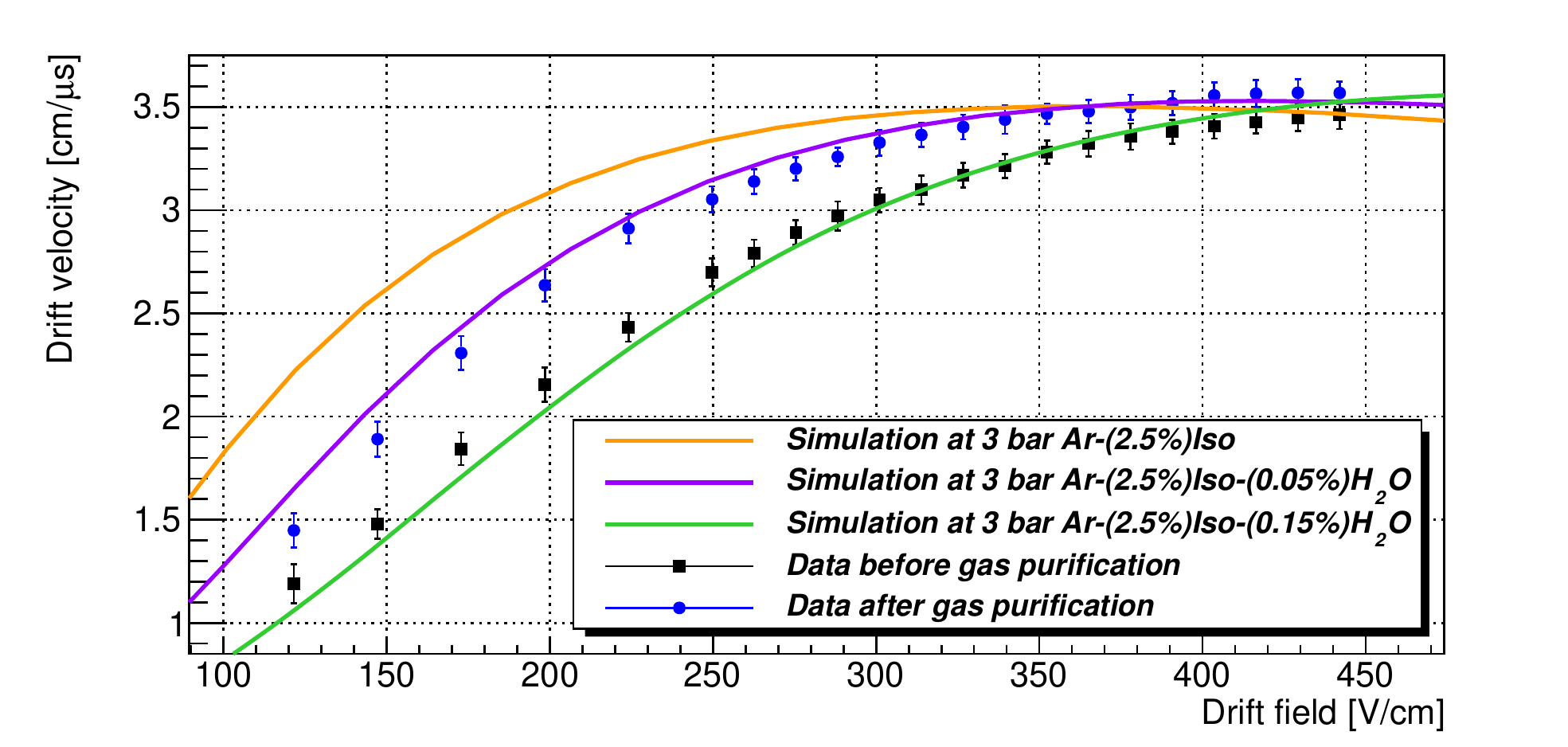}
    \caption{\label{fig:driftVelocity_vs_field} Dependence of the electron drift velocity on drift fields. The simulation lines, data before purification, and data after purification are all shown in the figure.}
\end{figure}

The drift fields were varied from 121 to 442 V/cm to analyze the impact on the drift velocity, as summarized in Fig.~\ref{fig:driftVelocity_vs_field}, while the amplification field was fixed at 52 kV/cm. 
Data were taken before gas purification (black points) and after gas purification (blue points), respectively. 
During the drift voltage scan, the drift velocity increased with the electric field until it plateaued at high fields. 
After gas purification, the electron drift velocity increased at each drift field, indicating an improvement in gas quality.
We also present three simulation lines from Garfield++ with different percentages of water contaminants to demonstrate the expected impact of impurities. 

\begin{figure}[htbp]
    \centering
    \includegraphics[width=0.6\textwidth]{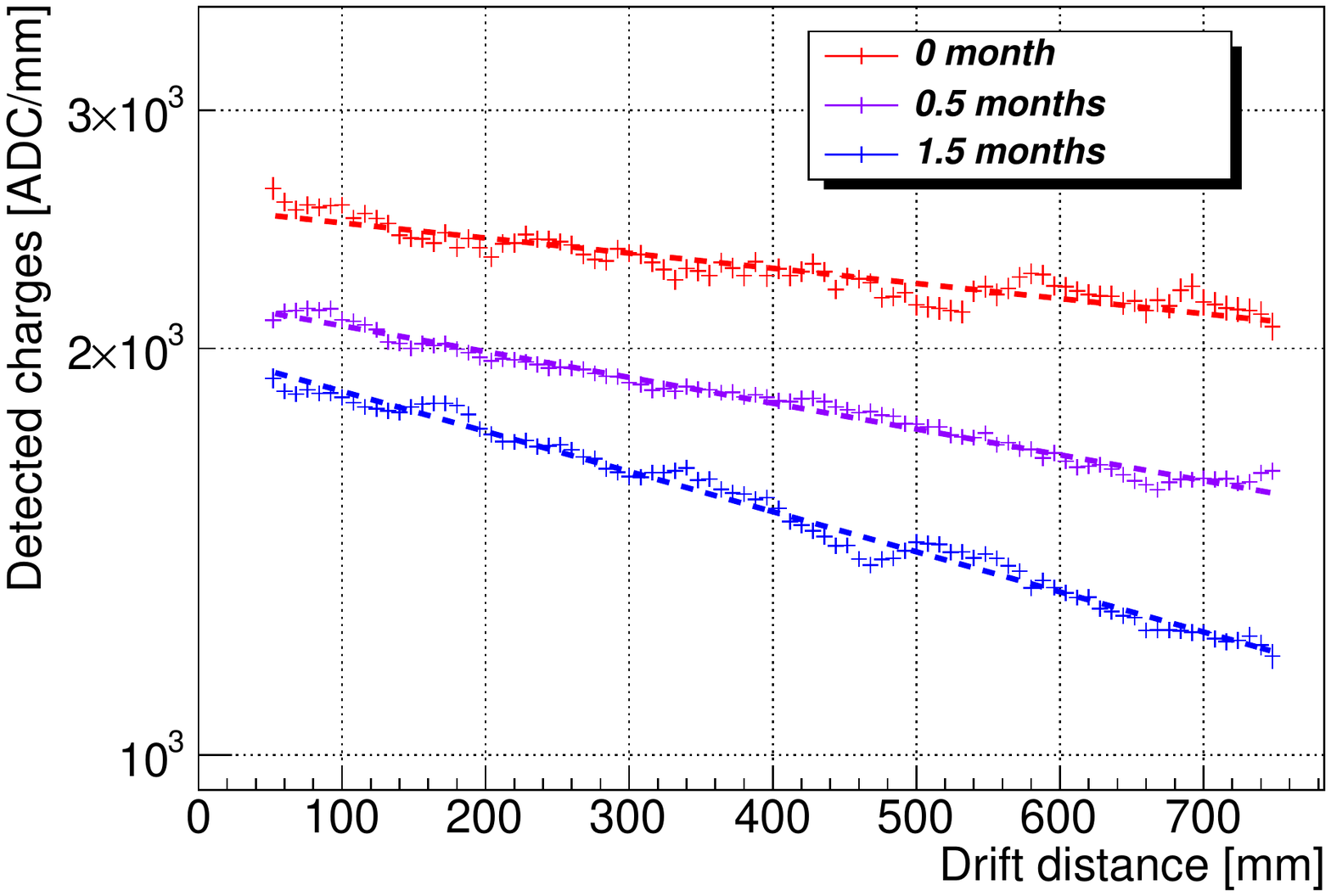}
    \caption{\label{fig:energyDeposition_vs_z} The average detected charges along the track as a function of the drift distance. The exponential fit was applied in the 50-750 mm range, as indicated by the dashed lines in the figure.}
\end{figure}

The mean energy loss rate dE/dx of high-energy muons in the TPC is almost constant~\cite{Groom:2001kq}.
Therefore, the differences in detected energy deposition along the through-going muon tracks from the anode to the cathode indicate the impact of the electron attachment effect and can be utilized to estimate electron absorption distance (lifetime). 
The energy deposition rate along the tracks can be calculated by summing the detected ADC samples of over-threshold signals, and then characterized in ADC/mm. 
Fig.~\ref{fig:energyDeposition_vs_z} shows the average detected charges along the track as a function of the drift distance.
Three sets of data at different times are presented.
On the first day of data taking, the electron absorption distance was ${\rm 237\pm 18 ~ cm}$ from the exponential fit.
The uncertainties are calculated with the fitting error for these measurements.
The value dwindles to ${\rm 173\pm 8 ~ cm}$ (${\rm 126\pm 4 ~ cm}$) after half a month (one and a half months) without gas purification.

\begin{figure}[htbp]
    \centering
    \includegraphics[width=0.9\textwidth]{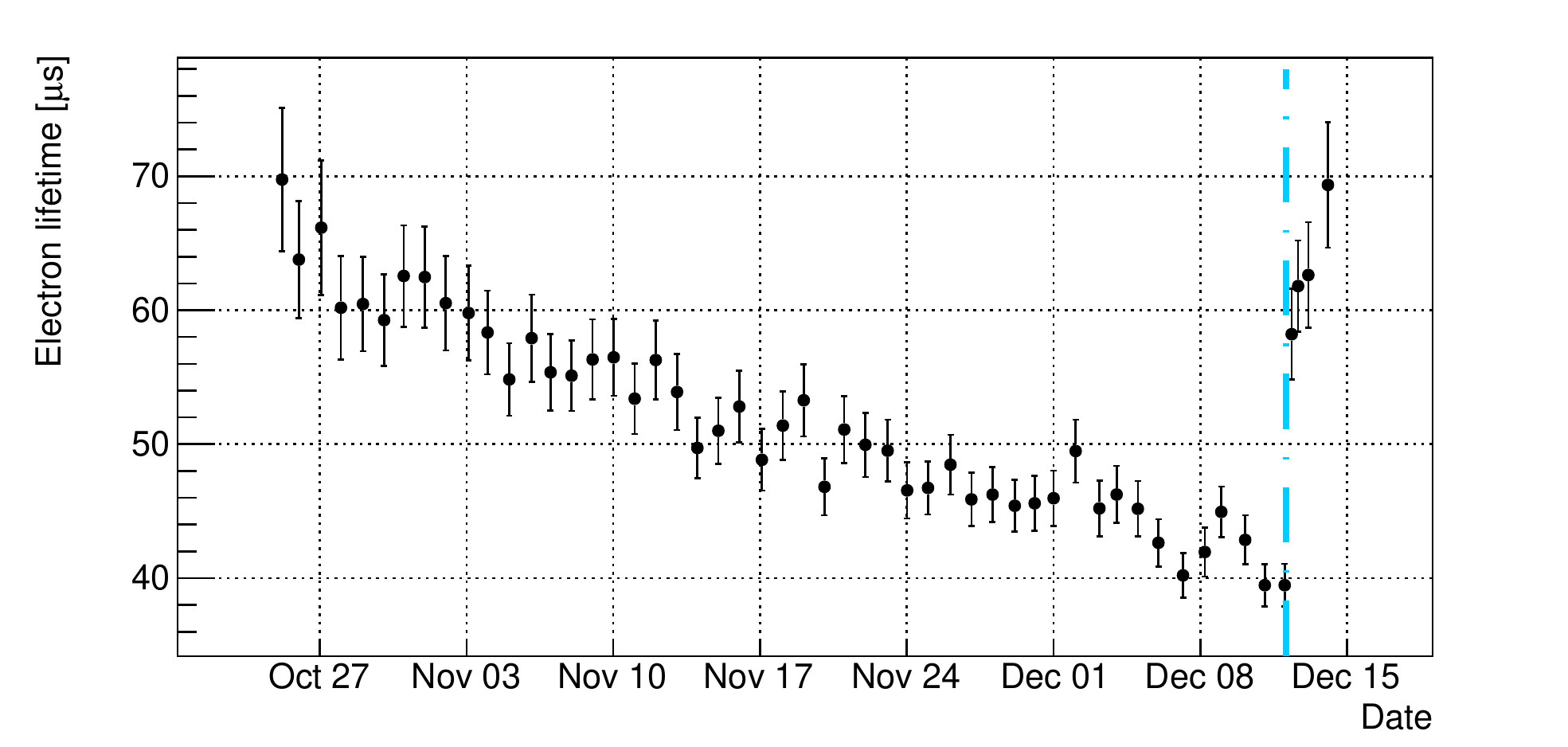}
    \caption{\label{fig:eLT_vs_time} The electron lifetime as a function of time. The blue dashed line denotes the starting time of the circulation and purification of the gas system. Data were taken at 3 bar Ar-(2.5\%)isobutane gas mixture and a drift field of 301 V/cm.}
\end{figure}

The evolution of electron lifetime calibrated with muons is shown in Fig.~\ref{fig:eLT_vs_time}. 
The electron lifetime decreased from ${\rm 69.8\pm 5.4 ~ \mu s}$ to ${\rm 39.5\pm 1.6 ~ \mu s}$ without gas purification in a long-term data-taking process.
The electron lifetime rapidly recovered with gas purification in two days.

\subsection{Detector gain}

\begin{figure}[htbp]
    \centering
    \includegraphics[width=0.6\textwidth]{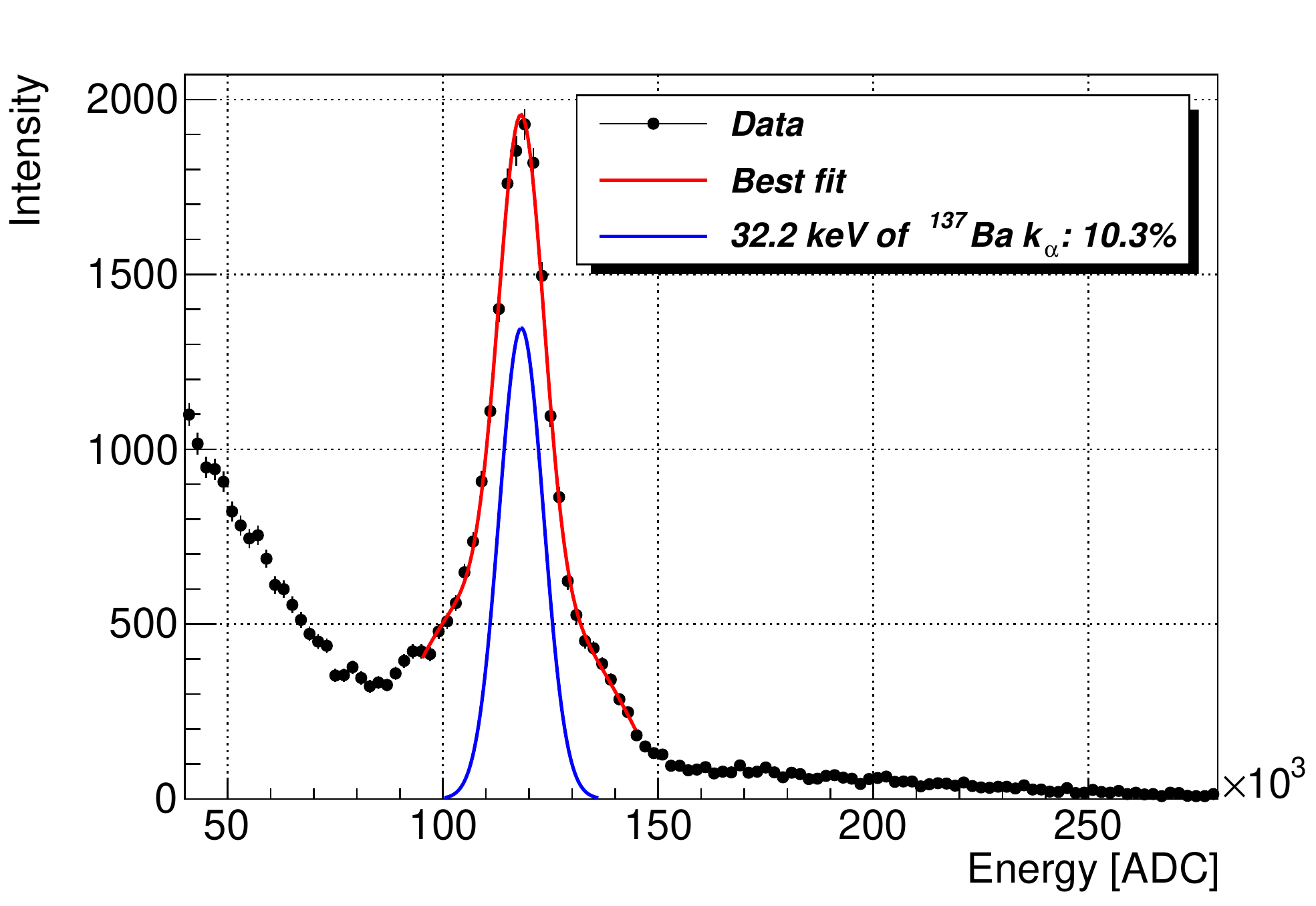}
    \caption{\label{fig:source_spectrum} The X-ray spectrum of the ${\rm ^{137}Cs}$ source with the positioning cut applied. }
\end{figure}

\begin{figure}[htbp]
    \centering
    \includegraphics[width=1.0\linewidth,trim=0 0 0 0,clip]{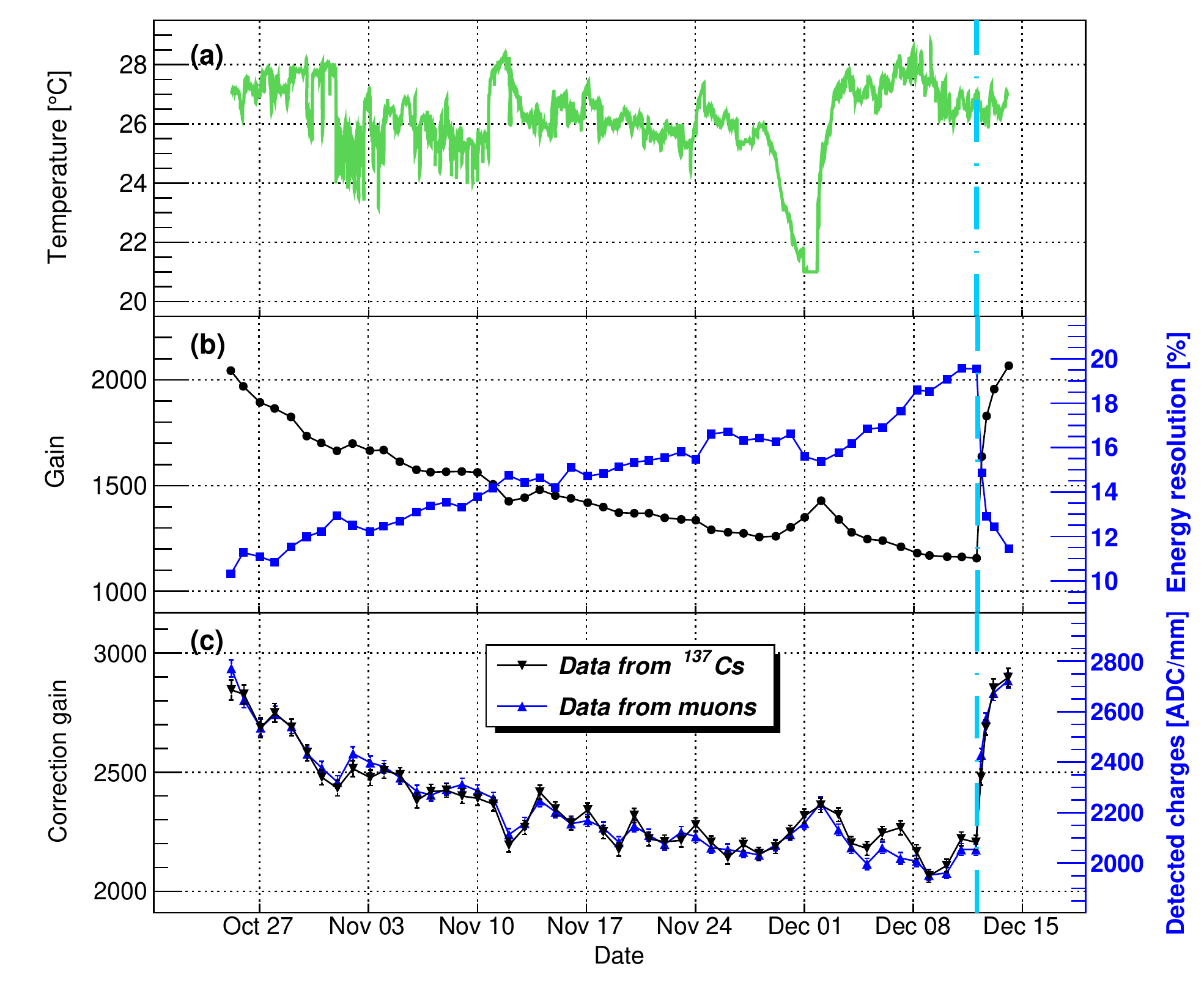}
    \caption{\label{fig:gain_vs_time} (a) Laboratory temperature. (b) Detector gain and Energy resolution at 32.2 keV from ${\rm ^{137}Cs}$. (c) Correction gain calibrated with ${\rm ^{137}Cs}$ (black points) and detected charges in ${z=0}$ characterized with through-going muons (blue points). The blue dashed line denotes the starting time of the circulation and purification of the gas system. Data were taken at 3 bar Ar-(2.5\%)isobutane gas mixture and a drift field of 301 V/cm.}
\end{figure}

The detector gain was calibrated with a $ ^{137}$Cs source.
It is defined as the ratio of the total charges of primary ionization electrons after and before the detector amplification:
\begin{equation}
    \label{eq:gain_calculation}
    G = \frac{Q}{(Ee)/\omega },
\end{equation}
where ${Q}$ represents the amplified charges collected by the electronics, ${E}$ represents the deposited energy of the corresponding incident particle, and ${e}$ is the elementary charge.  
The average ionization energy ${\omega}$ was taken as 26.2 eV for Ar-(2.5${\rm \%}$)Isobutane gas mixture \cite{Smirnov:2005yi}. 
The 32.2 keV X-ray peak from ${\rm ^{137}Cs}$, displayed in Fig.~\ref{fig:source_spectrum}, was fitted with a Gaussian plus a polynomial function to eliminate the linear background.
A gain of ${\rm 2043}$ and an energy resolution of ${\rm 10.3\%}$ FWHM (Full Width at Half Maximum) at 32.2 keV were obtained on the first day.
The evolution of gain and energy resolution at 32.2 keV are shown in Fig.~\ref{fig:gain_vs_time} (b). Gas quality dominated the evolution of energy resolution and detector gain in the TPC.
The laboratory temperature was also monitored to interpret gain curve fluctuations, as shown in Fig.~\ref{fig:gain_vs_time} (a). 
According to the gain model in~\cite{Iguaz:2012ur}, variations in the mean free path caused by the laboratory temperature affect the electron amplification process of the detector.

As mentioned in Equation~\ref{eq:electron_lifetime_calculation}, the detected charges were reduced due to electron absorption during drifting, and they can be corrected with ${z_e}$ estimated previously.
We set ${z}$ of ${\rm ^{137}Cs}$ signals to be 78.0~cm, as the location of ${\rm ^{137}Cs}$ source was on the cathode plane.
The correction gain calibrated with ${\rm ^{137}Cs}$ signals is presented in Fig.~\ref{fig:gain_vs_time} (c, black points).
The energy deposition rate of through-going muon events, extracted from the fit in Fig.~\ref{fig:energyDeposition_vs_z}, was also used to characterize the correction gain. 
The detected charges in ${z=0}$ characterized in ADC/mm are presented in Fig.~\ref{fig:gain_vs_time} (c, blue points). 
The tendency of correction gain calibrated with ${\rm ^{137}Cs}$ is consistent with that of detected charges in ${z=0}$ characterized with muons.
After correction, the gain evolution in principle was dominated by gas quality inside the Micromegas amplification gap.

\subsection{Electric field distortion}

The straightness of muon tracks provides an effective quantification of electric field distortion.
Distortion of the uniform electric field in the TPC active volume deforms the drift path of ionization electrons, diminishing the quality of reconstructed tracks. 
For a straight muon track, the deviation ${\rm \Delta X}$ (${\rm \Delta Y}$) between the detected and expected track can be obtained at certain Z positions in the XZ (YZ) plane with the aforementioned Hough transform. 
In this study, the track deviation is used to quantify the electric field distortion with the center Micromegas module for data taking (see Fig.~\ref{fig:readout_plane}). 

\begin{figure}[htbp]
    \centering
    \subfloat[]{
        \label{subfig:sim_field_comsol}
        \includegraphics[width=.52\textwidth,trim=130 300 130 300,clip]{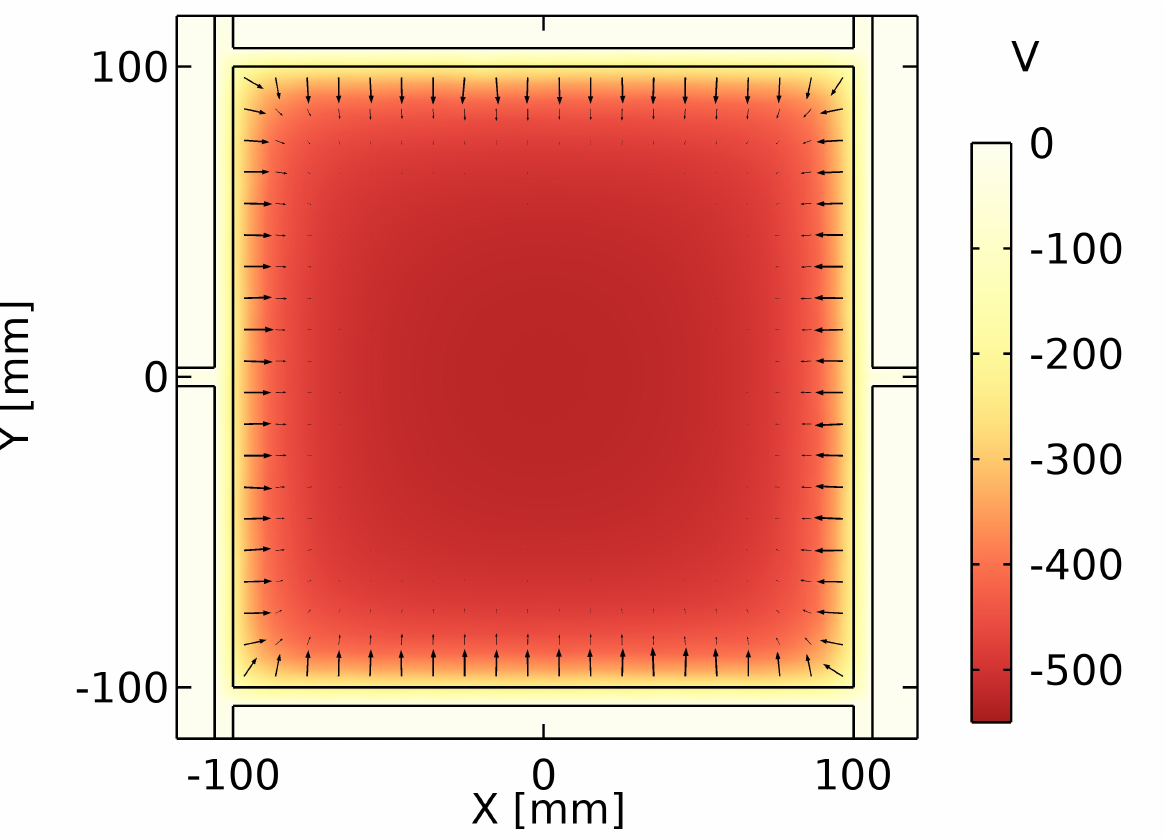}
        }
    \subfloat[]{
        \label{subfig:distortion_b}
        \includegraphics[width=.40\textwidth]{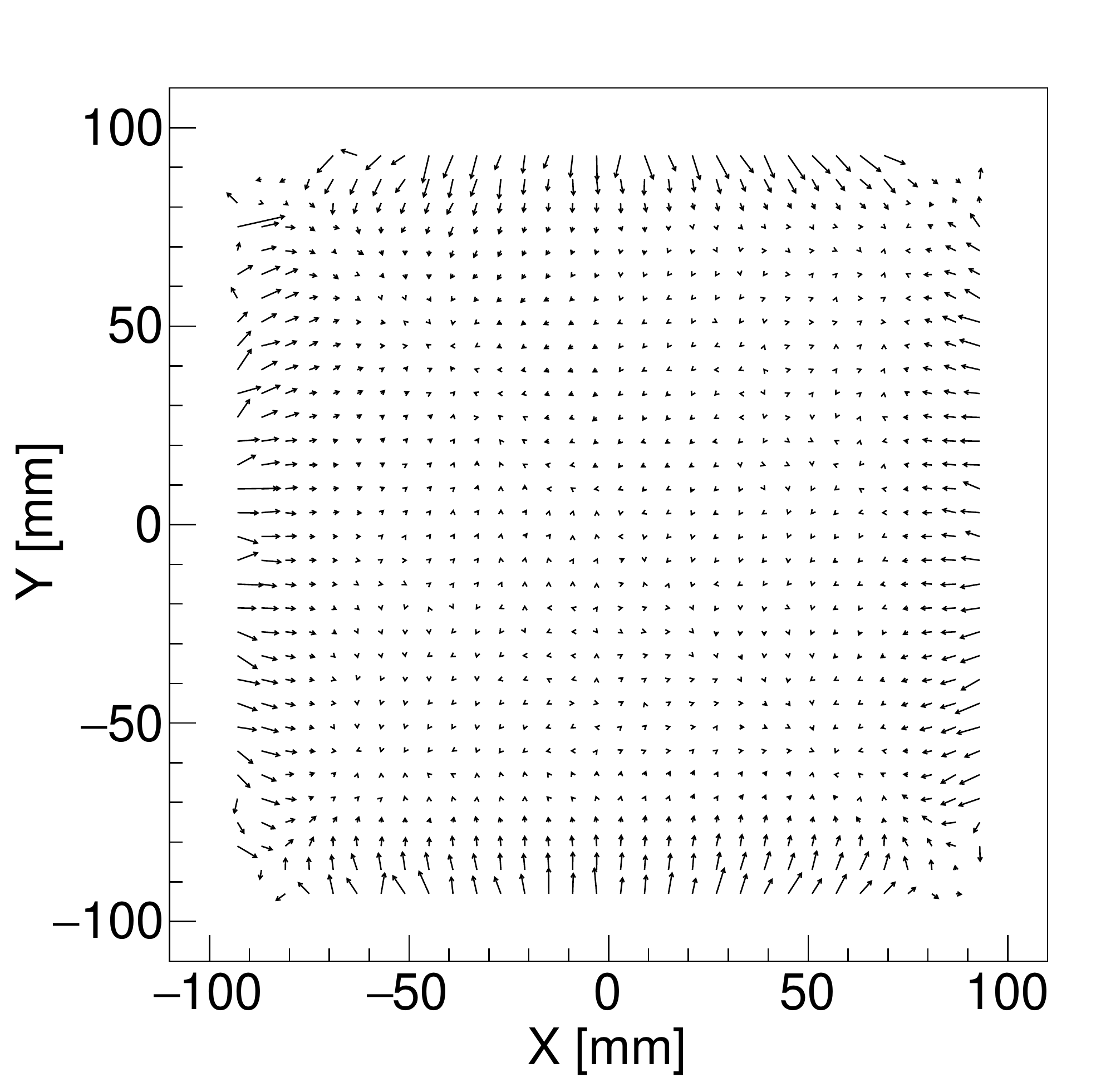}
    }
    \qquad
    \subfloat[]{
        \label{subfig:distortion_c}
        \includegraphics[width=.40\textwidth]{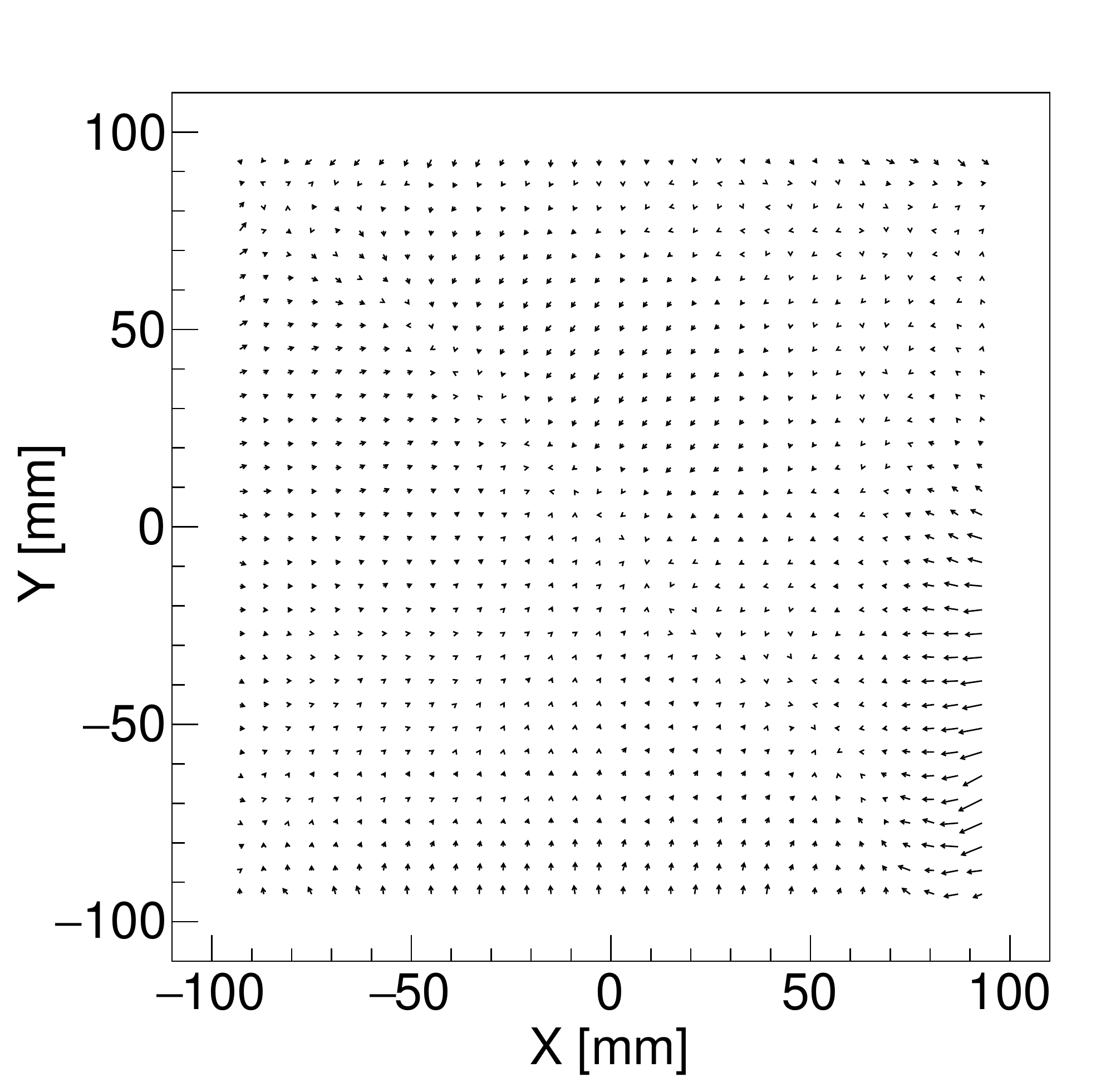}
    }
    \quad
    \quad
    \subfloat[]{
        \label{subfig:distortion_d}
        \includegraphics[width=.40\textwidth]{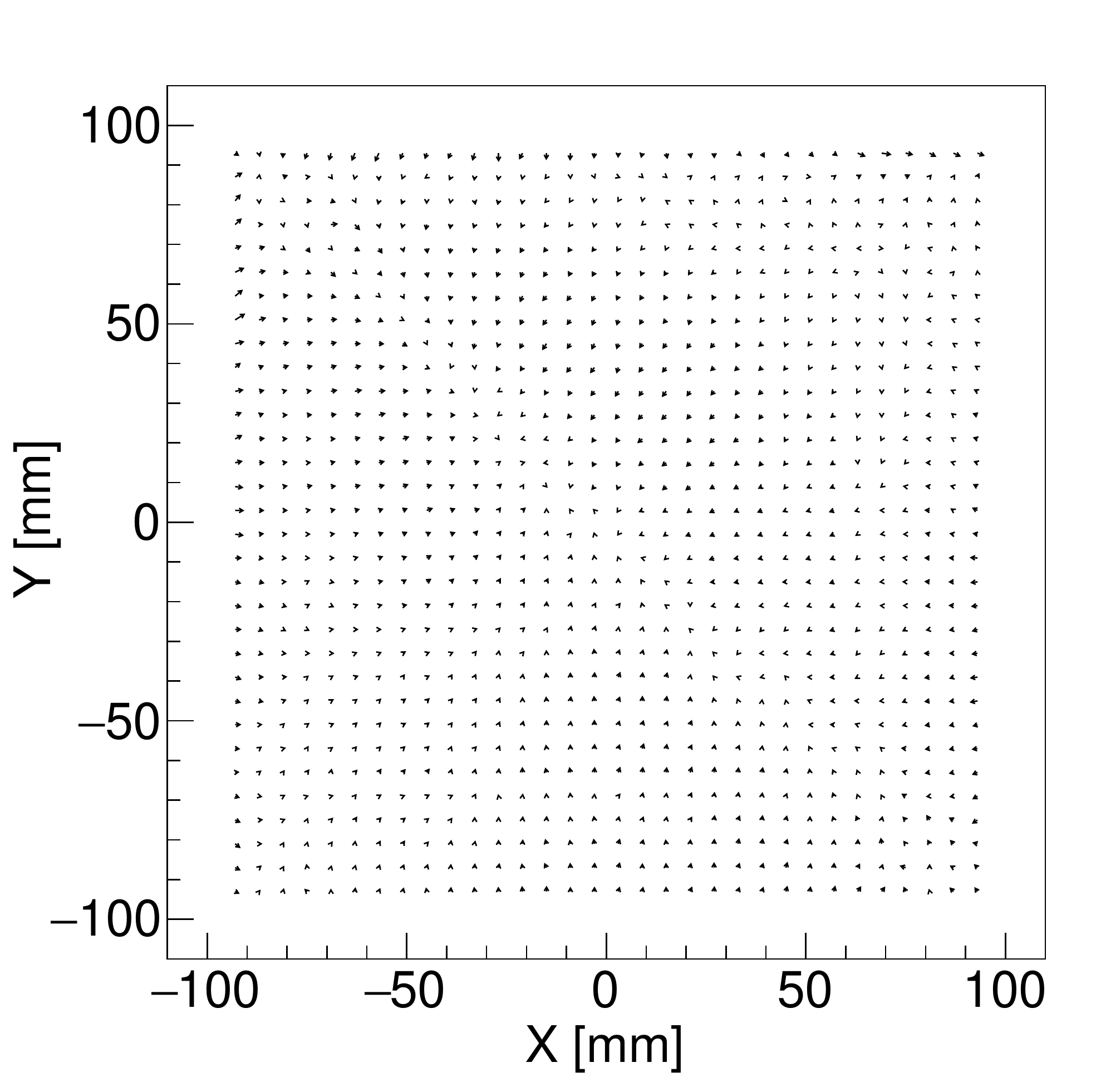}
    }
    \caption{\label{fig:distortion} 
    (a) The electric field map on the Micromegas plane simulated using COMSOL with only one data-taking Micromegas powered on. 
    (b) The electric field distortion plot derived from muon tracks with the same run conditions as the simulation. 
    (c) The electric field distortion plot with six Micromegas powered on, with the bottom right Micromegas turned off.
    (d) The electric field distortion plot with all seven Micromegas powered on.
    The arrows in these plots indicate the electric field direction and intensity on the data-taking Micromegas plane.}
\end{figure}

Electric field distortion often occurs at the Micromegas edge. 
All Micromegas modules were mounted on a grounded aluminum plane, and the electric potential difference at the Micromegas edge can cause electric field distortion. 
Additionally, if a high voltage was applied to one Micromegas module while the adjacent modules remained at zero voltage, distortion at the boundary was expected. 
Fig.~\ref{subfig:sim_field_comsol} shows a colored electric field map on the data-taking Micromegas module, simulated using COMSOL Multiphysics~\cite{COMSOL}. 
Only the center Micromegas was powered with a voltage of -520 V, while the other six were grounded. 
The arrows at the Micromegas edge illustrate the direction and intensity of field lines. 

The simulation results were confirmed with muon data. 
The Micromegas readout plane was divided into 32${\rm \times}$32 grids, and the average deviation of muon tracks in each grid was calculated to represent the electric field distortion. 
Fig.~\ref{subfig:distortion_b} shows an electric field distortion plot reconstructed with muon events under the same run conditions as the simulation. 
The distortion appeared near the boundary of Micromegas, and its direction is consistent with the simulation. 
Fig.~\ref{subfig:distortion_c} shows an electric field distortion plot with six Micromegas powered on at the same voltage. The Micromegas at the bottom-right corner was turned off, resulting in an electric field distortion at the bottom-right corner on the data-taking Micromegas. 
In Fig.~\ref{subfig:distortion_d}, all seven Micromegas modules were powered on, and a uniform electric field on the data-taking Micromegas is shown. 

\begin{figure}[htbp]
    \centering
    \includegraphics[width=0.9\textwidth]{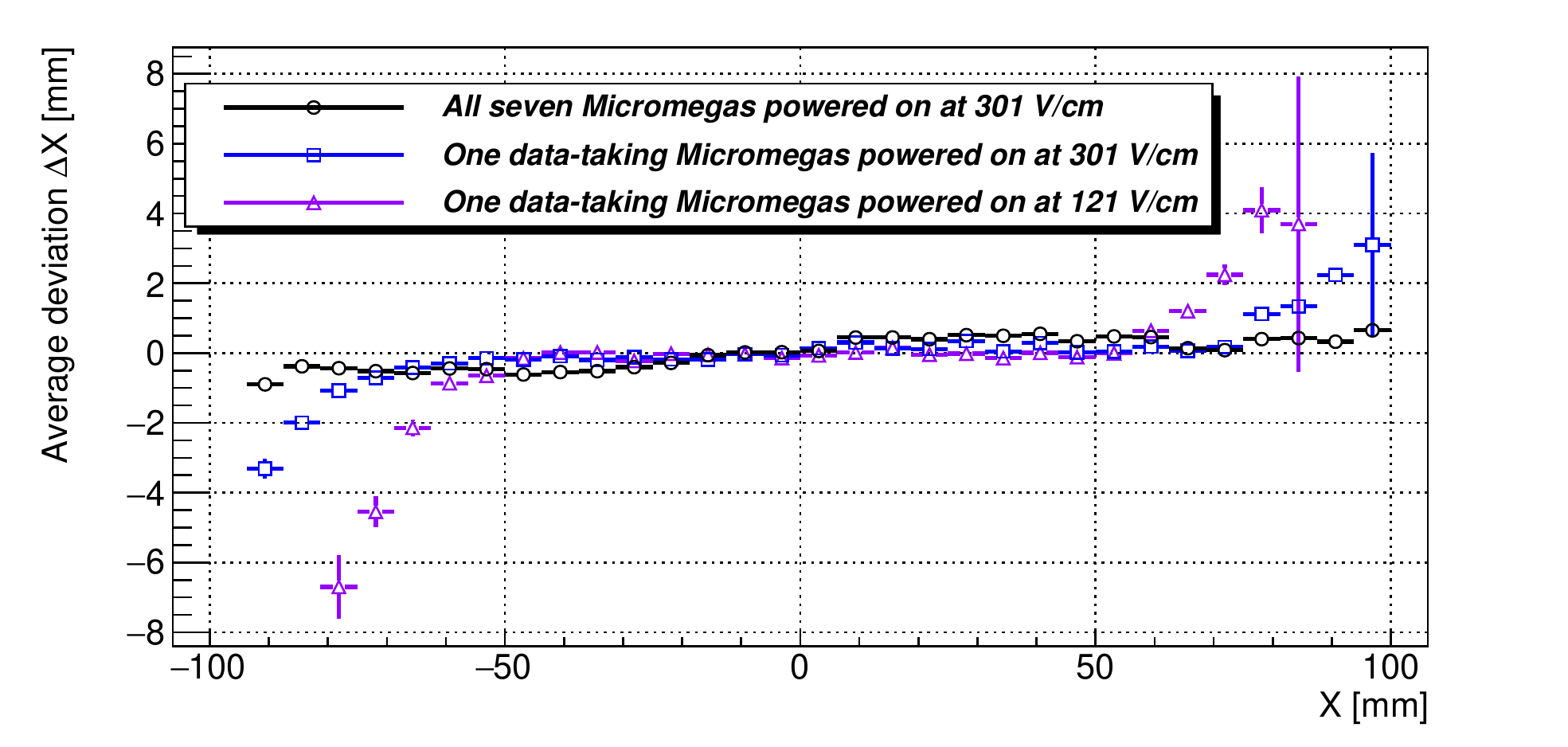}
    \caption{\label{fig:distortion_offset} The average deviation ${\rm \Delta X}$ of muon tracks evolves with X under different run conditions.}
\end{figure}

To quantify the distortion, we calculated the average deviation ${\rm \Delta X}$ as a function of the X position under different run conditions in Fig.~\ref{fig:distortion_offset}.
The X position distribution was divided into 32 segments, with each segment measuring 6 mm.
The Y position distribution was fixed between -3 and 3 mm. 
The expected muon track was reconstructed with the straight line from the Hough transform.
The deviation ${\rm \Delta X}$ was then calculated for each segment, comparing the detected track to the expected track. 
A negligible distortion with all seven Micromegas powered on and operated at a drift field of 301 V/cm was shown in the figure.
Under the same drift field, the distortion at the edge of the data-taking Micromegas module increased dramatically when the other six Micromegas modules were turned off.
Furthermore, the distortion became more severe at a low drift field of 121 V/cm, affecting an area of approximately 40 mm from the detector edge, with a maximum distortion deviation of more than 6 mm along the X direction. 
The distortion also resulted in a prominent dead zone, which was the reason for missing data points at the boundary.
The distortion analysis can provide a reference for the positioning cut in signal selection.

%% file: Conclusions.tex
\section{Conclusion and outlook}\label{sec:Conclusions}

In this paper, we report the calibration results of a high-pressure gaseous TPC with cosmic muon tracks. 
The TPC is read out by the Micromegas module, which has been recently developed and fabricated with the novel thermal bonding technique. 
The detector was operated continuously for 50 days with 3 bar Ar-(2.5${\rm \%}$)Isobutane gas mixture. 
The performances of the TPC, including the drift velocity, electron lifetime, detector gain, and electric field distortion, were characterized and calibrated using cosmic ray muons. 
The drift velocity (electron lifetime) degraded with gas deterioration from ${\rm 3.40\pm 0.07 ~ cm/\mu s}$ (${\rm 69.8\pm 5.4 ~ \mu s}$) to ${\rm 3.06\pm 0.06 ~ cm/\mu s}$ (${\rm 39.5\pm 1.6 ~ \mu s}$), and recovered with gas purification.
When the impact of electron lifetime is corrected, the tendency of detector gain calibrated with ${\rm ^{137}Cs}$ is consistent with that of detected charges characterized with muons.
Analyzing the electric field distortion under different run conditions also demonstrates the potential of straight muon tracks.

Cosmic ray muons provide a penetrating and non-hazardous calibration source for TPCs, which can be essential for detectors without the absolute $t_0$ information.
The obtained results are helpful for monitoring drift properties, gain stability, and gas purity of the detector of the PandaX-III experiment during commissioning.
The technique will also be applied to other gaseous TPCs, such as low-background alpha detectors~\cite{Du:2022elk}.
The method may be of interest to TPC detectors with $t_0$ for validation of timing measurements.
More systematic measurements and calibrations are in progress, especially with seven Micromegas readout modules and high-pressure xenon.
Further improvement of the muon track reconstruction algorithm, especially the precise measurement of incident angles, can help the future development of muon tomography~\cite{Bonomi:2020dmm}. 

%% file: PrototypeTPC_MuonCalibration.bbl
\begin{thebibliography}{10}

\bibitem{Marx:1978zz}
Jay~N. Marx and David~R. Nygren.
\newblock {The Time Projection Chamber}.
\newblock {\em Phys. Today}, 31N10:46--53, 1978.

\bibitem{NEXT:2015wlq}
J.~Mart\'\i{}n-Albo et~al.
\newblock {Sensitivity of NEXT-100 to Neutrinoless Double Beta Decay}.
\newblock {\em JHEP}, 05:159, 2016.

\bibitem{EXO-200:2019rkq}
G.~Anton et~al.
\newblock {Search for Neutrinoless Double-$\beta$ Decay with the Complete
  EXO-200 Dataset}.
\newblock {\em Phys. Rev. Lett.}, 123(16):161802, 2019.

\bibitem{XENON:2018voc}
E.~Aprile et~al.
\newblock {Dark Matter Search Results from a One Ton-Year Exposure of XENON1T}.
\newblock {\em Phys. Rev. Lett.}, 121(11):111302, 2018.

\bibitem{Iguaz:2015myh}
F.~J. Iguaz et~al.
\newblock {TREX-DM: a low-background Micromegas-based TPC for low-mass WIMP
  detection}.
\newblock {\em Eur. Phys. J. C}, 76(10):529, 2016.

\bibitem{PandaX-4T:2021bab}
Yue Meng et~al.
\newblock {Dark Matter Search Results from the PandaX-4T Commissioning Run}.
\newblock {\em Phys. Rev. Lett.}, 127(26):261802, 2021.

\bibitem{Gonzalez-Diaz:2017gxo}
D.~Gonzalez-Diaz, F.~Monrabal, and S.~Murphy.
\newblock {Gaseous and dual-phase time projection chambers for imaging rare
  processes}.
\newblock {\em Nucl. Instrum. Meth. A}, 878:200--255, 2018.

\bibitem{Li:2021viv}
Tao Li, Shaobo Wang, Yu~Chen, Ke~Han, Heng Lin, Kaixiang Ni, Wei Wang, Yiliu
  Xu, and An\textquoteright{}ni Zou.
\newblock {Signal identification with Kalman Filter towards background-free
  neutrinoless double beta decay searches in gaseous detectors}.
\newblock {\em JHEP}, 06:106, 2021.

\bibitem{Chen:2016qcd}
Xun Chen et~al.
\newblock {PandaX-III: Searching for neutrinoless double beta decay with high
  pressure$^{136}$Xe gas time projection chambers}.
\newblock {\em Sci. China Phys. Mech. Astron.}, 60(6):061011, 2017.

\bibitem{Giomataris:1995fq}
Y.~Giomataris, P.~Rebourgeard, J.~P. Robert, and Georges Charpak.
\newblock {MICROMEGAS: A High granularity position sensitive gaseous detector
  for high particle flux environments}.
\newblock {\em Nucl. Instrum. Meth. A}, 376:29--35, 1996.

\bibitem{Feng:2019prv}
Jianxin Feng, Zhiyong Zhang, Jianbei Liu, Binbin Qi, Anqi Wang, Ming Shao, and
  Yi~Zhou.
\newblock {A thermal bonding method for manufacturing Micromegas detectors}.
\newblock {\em Nucl. Instrum. Meth. A}, 989:164958, 2021.

\bibitem{Feng:2022jkd}
Jianxin Feng, Zhiyong Zhang, Jianbei Liu, Ming Shao, and Yi~Zhou.
\newblock {A novel resistive anode using a germanium film for Micromegas
  detectors}.
\newblock {\em Nucl. Instrum. Meth. A}, 1031:166595, 2022.

\bibitem{Iguaz:2012ur}
F.~J. Iguaz, E.~Ferrer-Ribas, A.~Giganon, and I.~Giomataris.
\newblock {Characterization of microbulk detectors in argon- and neon-based
  mixtures}.
\newblock {\em JINST}, 7:P04007, 2012.

\bibitem{Groom:2001kq}
Donald~E. Groom, Nikolai~V. Mokhov, and Sergei~I. Striganov.
\newblock {Muon stopping power and range tables 10-MeV to 100-TeV}.
\newblock {\em Atom. Data Nucl. Data Tabl.}, 78:183--356, 2001.

\bibitem{Kobayashi:2010hx}
M.~Kobayashi et~al.
\newblock {Cosmic ray tests of a GEM-based TPC prototype operated in
  Ar-CF4-isobutane gas mixtures}.
\newblock {\em Nucl. Instrum. Meth. A}, 641:37--47, 2011.
\newblock [Erratum: Nucl.Instrum.Meth.A 697, 122 (2013)].

\bibitem{WA105:2021zin}
B.~Aimard et~al.
\newblock {Performance study of a 3\texttimes{}1\texttimes{}1 m3 dual phase
  liquid Argon Time Projection Chamber exposed to cosmic rays}.
\newblock {\em JINST}, 16(08):P08063, 2021.

\bibitem{Wang:2020owr}
Shaobo Wang.
\newblock {The TPC detector of PandaX-III Neutrinoless Double Beta Decay
  experiment}.
\newblock {\em JINST}, 15(03):C03052, 2020.

\bibitem{Lin:2018mpd}
Heng Lin et~al.
\newblock {Design and commissioning of a 600 L Time Projection Chamber with
  Microbulk Micromegas}.
\newblock {\em JINST}, 13(06):P06012, 2018.

\bibitem{Giovinazzo:2016ikh}
J.~Giovinazzo et~al.
\newblock {GET electronics samples data analysis}.
\newblock {\em Nucl. Instrum. Meth. A}, 840:15--27, 2016.

\bibitem{Anvar2011AGETTG}
S.~Anvar et~al.
\newblock Aget, the get front-end asic, for the readout of the time projection
  chambers used in nuclear physic experiments.
\newblock {\em 2011 IEEE Nuclear Science Symposium Conference Record}, pages
  745--749, 2011.

\bibitem{NEXT:2018wtg}
L.~Rogers et~al.
\newblock {High Voltage Insulation and Gas Absorption of Polymers in High
  Pressure Argon and Xenon Gases}.
\newblock {\em JINST}, 13(10):P10002, 2018.

\bibitem{Lin:2022kua}
Heng Lin et~al.
\newblock {Measurement of high-pressure xenon gas absorption in acrylic}.
\newblock {\em JINST}, 17(05):P05027, 2022.

\bibitem{SAES}
SAES~Getters Group.
\newblock \url{https://www.saesgetters.com/}.

\bibitem{Duda:1972ymn}
Richard~O. Duda and Peter~E. Hart.
\newblock {Use of the Hough transformation to detect lines and curves in
  pictures}.
\newblock {\em Commun. ACM}, 15(1):11--15, 1972.

\bibitem{Altenmuller:2021slh}
Konrad Altenm\"uller et~al.
\newblock {REST-for-Physics, a ROOT-based framework for event oriented data
  analysis and combined Monte Carlo response}.
\newblock {\em Comput. Phys. Commun.}, 273:108281, 2022.

\bibitem{Huxley:1974}
L.G.H. Huxley and R.W. Crompton.
\newblock {\em Diffusion and drift of electrons in gases}.
\newblock Wiley, 1974.

\bibitem{Garfield++}
R.~Veenhof.
\newblock Simulation of gaseous detectors.
\newblock \url{https://garfield.web.cern.ch/garfield/}.

\bibitem{matweb}
\url{https://www.matweb.com/Search/MaterialGroupSearch.aspx?GroupID=63}.

\bibitem{Smirnov:2005yi}
I.~B. Smirnov.
\newblock {Modeling of ionization produced by fast charged particles in gases}.
\newblock {\em Nucl. Instrum. Meth. A}, 554:474--493, 2005.

\bibitem{COMSOL}
COMSOL Multiphysics~V. 6.1.
\newblock Comsol {AB}, {S}tockholm, {S}weden.
\newblock \url{https://www.comsol.com}.

\bibitem{Du:2022elk}
Haiyan Du et~al.
\newblock {A gaseous time projection chamber with Micromegas readout for
  low-radioactive material screening}.
\newblock {\em Radiat. Detect. Technol. Methods}, 7(1):90--99, 2023.

\bibitem{Bonomi:2020dmm}
G.~Bonomi, P.~Checchia, M.~D'Errico, D.~Pagano, and G.~Saracino.
\newblock {Applications of cosmic-ray muons}.
\newblock {\em Prog. Part. Nucl. Phys.}, 112:103768, 2020.

\end{thebibliography}
